\pdfoutput=1

\documentclass[12pt,a4paper]{article}

\usepackage{ifthen} 
\newboolean{pdflatex}
\setboolean{pdflatex}{true} 

\newboolean{articletitles}
\setboolean{articletitles}{true} 

\newboolean{uprightparticles}
\setboolean{uprightparticles}{false} 

\usepackage{microtype}
\usepackage{lineno}  
\usepackage{xspace} 

\usepackage{graphicx}  
\usepackage{color}
\usepackage{colortbl}
\graphicspath{{./figs/}} 

\usepackage{amsmath} 
\usepackage{amssymb}
\usepackage{amsfonts}
\usepackage{upgreek} 

\newcommand*\patchAmsMathEnvironmentForLineno[1]{%
\expandafter\let\csname old#1\expandafter\endcsname\csname #1\endcsname
\expandafter\let\csname oldend#1\expandafter\endcsname\csname
end#1\endcsname
 \renewenvironment{#1}%
   {\linenomath\csname old#1\endcsname}%
   {\csname oldend#1\endcsname\endlinenomath}%
}
\newcommand*\patchBothAmsMathEnvironmentsForLineno[1]{%
  \patchAmsMathEnvironmentForLineno{#1}%
  \patchAmsMathEnvironmentForLineno{#1*}%
}
\AtBeginDocument{%
\patchBothAmsMathEnvironmentsForLineno{equation}%
\patchBothAmsMathEnvironmentsForLineno{eqnarray}%
\patchBothAmsMathEnvironmentsForLineno{align}%
\patchBothAmsMathEnvironmentsForLineno{flalign}%
\patchBothAmsMathEnvironmentsForLineno{alignat}%
\patchBothAmsMathEnvironmentsForLineno{gather}%
\patchBothAmsMathEnvironmentsForLineno{multline}%
}

\usepackage{hyperref}    
\usepackage[all]{hypcap} 

\def\lhcb {\mbox{LHCb}\xspace}
\def\ux85 {\mbox{UX85}\xspace}

\def\babar  {\mbox{BaBar}\xspace}
\def\belle  {\mbox{Belle}\xspace}

\def\cdf    {\mbox{CDF}\xspace}

\ifthenelse{\boolean{uprightparticles}}%
{

 \def\Pmu         {\ensuremath{\upmu}\xspace}

 \def\Ppi         {\ensuremath{\uppi}\xspace}

 \def\Ptau        {\ensuremath{\uptau}\xspace}

 \def\Ppsi        {\ensuremath{\uppsi}\xspace}

 \def\PDelta      {\ensuremath{\Delta}\xspace}                 
 \def\PXi      {\ensuremath{\Xi}\xspace}                 
 \def\PLambda      {\ensuremath{\Lambda}\xspace}                 
 \def\PSigma      {\ensuremath{\Sigma}\xspace}                 
 \def\POmega      {\ensuremath{\Omega}\xspace}                 
 \def\PUpsilon      {\ensuremath{\Upsilon}\xspace}                 
 
 \def\PB      {\ensuremath{\mathrm{B}}\xspace}                 
                  
 \def\PD      {\ensuremath{\mathrm{D}}\xspace}

 \def\PJ      {\ensuremath{\mathrm{J}}\xspace}                 
 \def\PK      {\ensuremath{\mathrm{K}}\xspace}

 \def\PX      {\ensuremath{\mathrm{X}}\xspace}

 \def\Pb      {\ensuremath{\mathrm{b}}\xspace}                 
 \def\Pc      {\ensuremath{\mathrm{c}}\xspace}

 \def\Pi      {\ensuremath{\mathrm{i}}\xspace}

 \def\Pp      {\ensuremath{\mathrm{p}}\xspace}

 \def\Ps      {\ensuremath{\mathrm{s}}\xspace}

}
{

 \def\Pmu         {\ensuremath{\mu}\xspace}

 \def\Ppi         {\ensuremath{\pi}\xspace}

 \def\Ptau        {\ensuremath{\tau}\xspace}

 \def\Ppsi        {\ensuremath{\psi}\xspace}                 
                  
 \mathchardef\PDelta="7101
 \mathchardef\PXi="7104
 \mathchardef\PLambda="7103
 \mathchardef\PSigma="7106
 \mathchardef\POmega="710A
 \mathchardef\PUpsilon="7107
                  
 \def\PB      {\ensuremath{B}\xspace}                 
                  
 \def\PD      {\ensuremath{D}\xspace}

 \def\PJ      {\ensuremath{J}\xspace}                 
 \def\PK      {\ensuremath{K}\xspace}

 \def\PX      {\ensuremath{X}\xspace}

 \def\Pb      {\ensuremath{b}\xspace}                 
 \def\Pc      {\ensuremath{c}\xspace}

 \def\Pi      {\ensuremath{i}\xspace}

 \def\Pp      {\ensuremath{p}\xspace}

 \def\Ps      {\ensuremath{s}\xspace}

}

\def\mup        {\ensuremath{\Pmu^+}\xspace}
\def\mun        {\ensuremath{\Pmu^-}\xspace} 

\def\squark    {\ensuremath{\Ps}\xspace}

\def\cquark    {\ensuremath{\Pc}\xspace}

\def\bquark    {\ensuremath{\Pb}\xspace}
\def\bquarkbar {\ensuremath{\overline \bquark}\xspace}

\def\pion  {\ensuremath{\Ppi}\xspace}
\def\piz   {\ensuremath{\pion^0}\xspace}

\def\pip   {\ensuremath{\pion^+}\xspace}
\def\pim   {\ensuremath{\pion^-}\xspace}

\def\kaon  {\ensuremath{\PK}\xspace}
  \def\Kbar  {\kern 0.2em\overline{\kern -0.2em \PK}{}\xspace}

\def\Kz    {\ensuremath{\kaon^0}\xspace}
\def\Kzb   {\ensuremath{\Kbar^0}\xspace}
\def\KzKzb {\ensuremath{\Kz \kern -0.16em \Kzb}\xspace}
\def\Kp    {\ensuremath{\kaon^+}\xspace}
\def\Km    {\ensuremath{\kaon^-}\xspace}

\def\KpKm  {\ensuremath{\Kp \kern -0.16em \Km}\xspace}

  \def\Dbar    {\kern 0.2em\overline{\kern -0.2em \PD}{}\xspace}
\def\D       {\ensuremath{\PD}\xspace}

\def\Dz      {\ensuremath{\D^0}\xspace}
\def\Dzb     {\ensuremath{\Dbar^0}\xspace}
\def\DzDzb   {\ensuremath{\Dz {\kern -0.16em \Dzb}}\xspace}
\def\Dp      {\ensuremath{\D^+}\xspace}
\def\Dm      {\ensuremath{\D^-}\xspace}

\def\DpDm    {\ensuremath{\Dp {\kern -0.16em \Dm}}\xspace}

\def\Dstarp  {\ensuremath{\D^{*+}}\xspace}

\def\Dstarpm {\ensuremath{\D^{*\pm}}\xspace}

\def\B       {\ensuremath{\PB}\xspace}
\def\Bbar    {\ensuremath{\kern 0.18em\overline{\kern -0.18em \PB}{}}\xspace}

\def\Bz      {\ensuremath{\B^0}\xspace}

\def\Bu      {\ensuremath{\B^+}\xspace}

\def\Bp      {\ensuremath{\Bu}\xspace}

\def\Bd      {\ensuremath{\B^0}\xspace}
\def\Bs      {\ensuremath{\B^0_\squark}\xspace}

\def\jpsi     {\ensuremath{{\PJ\mskip -3mu/\mskip -2mu\Ppsi\mskip 2mu}}\xspace}
\def\psitwos  {\ensuremath{\Ppsi{(2S)}}\xspace}

  \def\Y#1S{\ensuremath{\PUpsilon{(#1S)}}\xspace}

\def\proton      {\ensuremath{\Pp}\xspace}

\def\L {\ensuremath{\PLambda}\xspace}
\def\Lbar {\ensuremath{\kern 0.1em\overline{\kern -0.1em\PLambda}}\xspace}

\def\Lb      {\ensuremath{\L^0_\bquark}\xspace}

\def\Lc      {\ensuremath{\L^+_\cquark}\xspace}

\newcommand{\decay}[2]{\ensuremath{#1\!\to #2}\xspace}         

\def\to                 {\ensuremath{\rightarrow}\xspace}

\def\order   {\ensuremath{\mathcal{O}}\xspace}

\def\CP                {\ensuremath{C\!P}\xspace}

\newcommand{\mistag}{\ensuremath{\omega}\xspace}

\def\AT#1     {\ensuremath{A_{\mathrm{T}}^{#1}}\xspace}           

\def\C#1      {\ensuremath{\mathcal{C}_{#1}}\xspace}                       
\def\Cp#1     {\ensuremath{\mathcal{C}_{#1}^{'}}\xspace}                    
\def\Ceff#1   {\ensuremath{\mathcal{C}_{#1}^{\mathrm{(eff)}}}\xspace}        
\def\Cpeff#1  {\ensuremath{\mathcal{C}_{#1}^{'\mathrm{(eff)}}}\xspace}       
\def\Ope#1    {\ensuremath{\mathcal{O}_{#1}}\xspace}                       
\def\Opep#1   {\ensuremath{\mathcal{O}_{#1}^{'}}\xspace}                    

\def\dkpi       {\decay{\PD}{\PK\Ppi}}
\def\dkk        {\decay{\PD}{\PK\PK}}

\newcommand{\tev}{\ensuremath{\mathrm{\,Te\kern -0.1em V}}\xspace}
\newcommand{\gev}{\ensuremath{\mathrm{\,Ge\kern -0.1em V}}\xspace}
\newcommand{\mev}{\ensuremath{\mathrm{\,Me\kern -0.1em V}}\xspace}
\newcommand{\kev}{\ensuremath{\mathrm{\,ke\kern -0.1em V}}\xspace}
\newcommand{\ev}{\ensuremath{\mathrm{\,e\kern -0.1em V}}\xspace}
\newcommand{\gevc}{\ensuremath{{\mathrm{\,Ge\kern -0.1em V\!/}c}}\xspace}
\newcommand{\mevc}{\ensuremath{{\mathrm{\,Me\kern -0.1em V\!/}c}}\xspace}
\newcommand{\gevcc}{\ensuremath{{\mathrm{\,Ge\kern -0.1em V\!/}c^2}}\xspace}
\newcommand{\gevgevcccc}{\ensuremath{{\mathrm{\,Ge\kern -0.1em V^2\!/}c^4}}\xspace}
\newcommand{\mevcc}{\ensuremath{{\mathrm{\,Me\kern -0.1em V\!/}c^2}}\xspace}

\def\mum  {\ensuremath{\,\upmu\rm m}\xspace}

\def\invfb   {\ensuremath{\mbox{\,fb}^{-1}}\xspace}

\def\fs   {\ensuremath{\rm \,fs}\xspace}

\newcommand{\stat}{\ensuremath{\mathrm{\,(stat)}}\xspace}
\newcommand{\syst}{\ensuremath{\mathrm{\,(syst)}}\xspace}

\def\order{{\ensuremath{\cal O}}\xspace}

\def\gsim{{~\raise.15em\hbox{$>$}\kern-.85em
          \lower.35em\hbox{$\sim$}~}\xspace}
\def\lsim{{~\raise.15em\hbox{$<$}\kern-.85em
          \lower.35em\hbox{$\sim$}~}\xspace}

\newcommand{\mean}[1]{\ensuremath{\langle #1 \rangle}} 

\def\sPlot{\mbox{\em sPlot}}

\def\pt         {\mbox{$p_{\rm T}$}\xspace}

\def\dllkpi     {\ensuremath{\mathrm{DLL}_{\kaon\pion}}\xspace}

\def\mphi       {\mbox{$\phi$}\xspace}

\def\evtgen     {\mbox{\textsc{EvtGen}}\xspace}
\def\pythia     {\mbox{\textsc{Pythia}}\xspace}

\def\geant      {\mbox{\textsc{Geant4}}\xspace}

\def\photos     {\mbox{\textsc{Photos}}\xspace}

\def\tell1  {TELL1\xspace}
\def\ukl1   {UKL1\xspace}

\newcommand{\eg}{\mbox{e.g.}\xspace}
\newcommand{\ie}{\mbox{i.e.}}

\newcommand{\deltaACP}{\ensuremath{\Delta A_{\CP}}\xspace}
\newcommand{\AcpKK}{\ensuremath{A_{\CP}(\Km\Kp)}\xspace}
\newcommand{\Acppipi}{\ensuremath{A_{\CP}(\pim\pip)}\xspace}
\newcommand{\ACP}{\ensuremath{A_{\CP}}\xspace}

\newcommand{\AKK}{\ensuremath{A_{\rm raw}(\Km\Kp)}\xspace}
\newcommand{\Apipi}{\ensuremath{A_{\rm raw}(\pim\pip)}\xspace}
\newcommand{\AKpiUnw}{\ensuremath{A_{\rm raw}^{\rm unweighted}(\Km\pip)}\xspace}
\newcommand{\AKKUnw}{\ensuremath{A_{\rm raw}^{\rm unweighted}(\Km\Kp)}\xspace}
\newcommand{\ApipiUnw}{\ensuremath{A_{\rm raw}^{\rm unweighted}(\pim\pip)}\xspace}
\newcommand{\Araw}{\ensuremath{A_{\rm raw}}\xspace}

\newcommand{\DACP}{\ensuremath{\Delta A_{\CP}}\xspace}
\newcommand{\DACPraw}{\ensuremath{\deltaACP}\xspace}
\newcommand{\DACPUnw}{\ensuremath{\deltaACP^{\rm unweighted}}\xspace}

\newcommand{\AP}{\ensuremath{A_P^{\PB}}\xspace}
\newcommand{\AD}{\ensuremath{A_D^\mu}\xspace}

\newcommand{\mistagP}{\ensuremath{\mistag^+}\xspace}
\newcommand{\mistagM}{\ensuremath{\mistag^-}\xspace}

\def\dkk        {\decay{\Dz}{\Km\Kp}}
\def\dpipi      {\decay{\Dz}{\pim\pip}}
\def\dkpi       {\decay{\Dz}{\Km\pip}}

\usepackage{mciteplus}

\usepackage{longtable} 

\begin{document}

\renewcommand{\thefootnote}{\fnsymbol{footnote}}
\setcounter{footnote}{1}


\begin{titlepage}
\pagenumbering{roman}

\vspace*{-1.5cm}
\centerline{\large EUROPEAN ORGANIZATION FOR NUCLEAR RESEARCH (CERN)}
\vspace*{1.5cm}
\hspace*{-0.5cm}
\begin{tabular*}{\linewidth}{lc@{\extracolsep{\fill}}r}
\ifthenelse{\boolean{pdflatex}}
{\vspace*{-2.7cm}\mbox{\!\!\!\includegraphics[width=.14\textwidth]{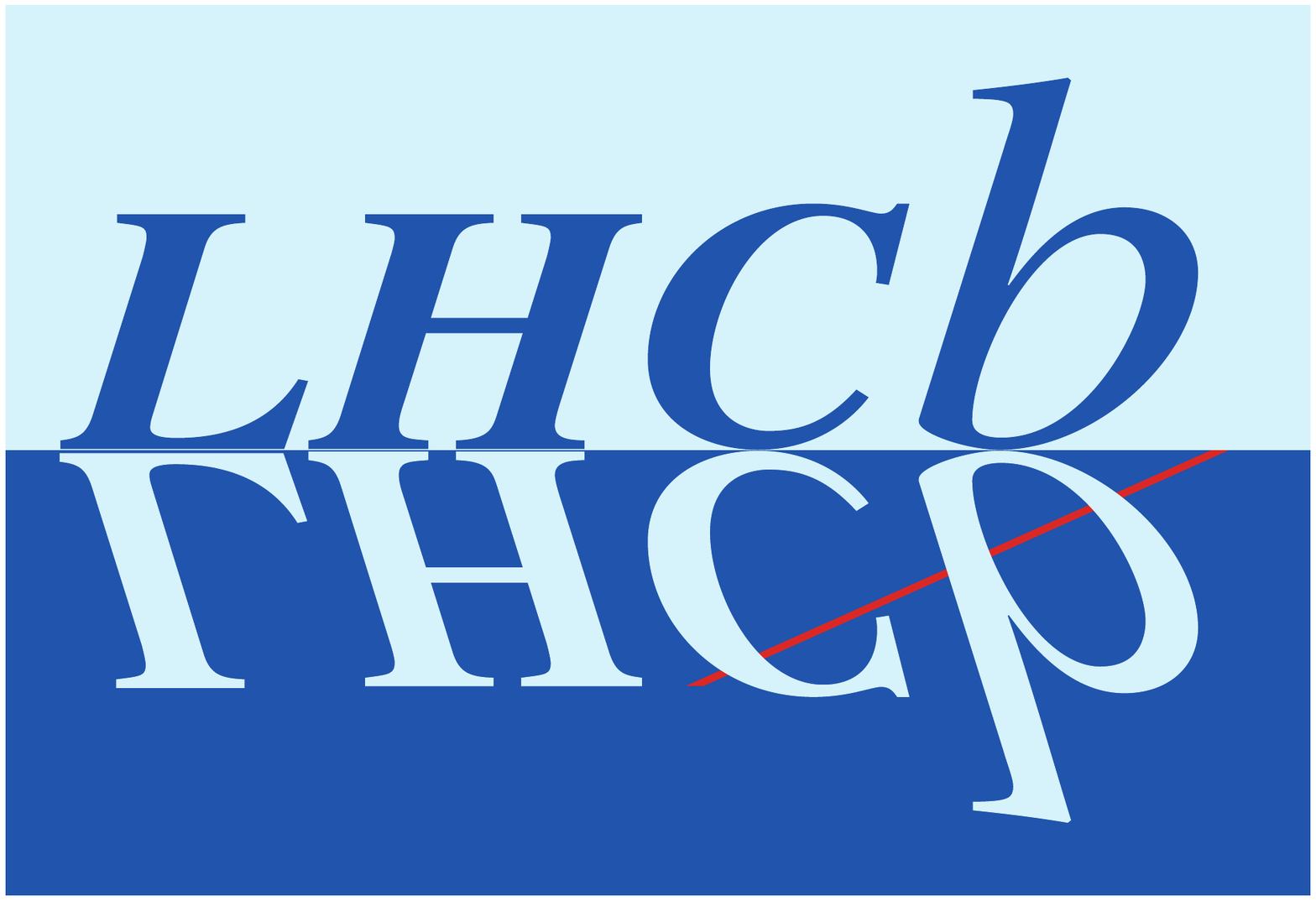}} & &}%
{\vspace*{-1.2cm}\mbox{\!\!\!\includegraphics[width=.12\textwidth]{lhcb-logo.eps}} & &}%
\\
 & & CERN-PH-EP-2013-039 \\  
 & & LHCb-PAPER-2013-003 \\  
 & & 29 April 2013 \\ 
 & & \\
\end{tabular*}

\vspace*{2.5cm}

{\bf\boldmath\huge
\begin{center}
  Search for direct \CP violation in $D^0 \to h^- h^+$ modes using semileptonic $B$ decays
\end{center}
}

\vspace*{1.5cm}

\begin{center}
The LHCb collaboration\footnote{Authors are listed on the following pages.}
\end{center}

\vspace{\fill}

\begin{abstract}
  \noindent
A search for direct \CP violation in $\Dz\to h^-h^+$ (where $h=K$ or $\pi$) is presented using data corresponding to an integrated luminosity of
$1.0\invfb$ collected in 2011 by LHCb in $pp$ collisions at a centre-of-mass
energy of 7\tev. The analysis uses \Dz mesons produced in inclusive semileptonic
\bquark-hadron decays to the $\Dz\mu\PX$ final state, where the charge of the
accompanying muon is used to tag the flavour of the \Dz meson. The difference in
the \CP-violating asymmetries between the two decay channels is measured to be
\begin{equation}
   \DACP = \AcpKK-\Acppipi = (0.49\pm 0.30\stat \pm 0.14\syst)\%
   \ . \nonumber
\end{equation}
This result does not confirm the evidence for direct \CP violation in the charm
sector reported in other analyses.
\end{abstract}

\vspace*{1.5cm}

\begin{center}
  Published in Phys. Lett. B
\end{center}

\vspace{\fill}

{\footnotesize
\centerline{\copyright~CERN on behalf of the \lhcb collaboration, license \href{http://creativecommons.org/licenses/by/3.0/}{CC-BY-3.0}.}}
\vspace*{2mm}

\end{titlepage}



\newpage
\setcounter{page}{2}
\mbox{~}
\newpage

\centerline{\large\bf LHCb collaboration}
\begin{flushleft}
\small
R.~Aaij$^{40}$, 
C.~Abellan~Beteta$^{35,n}$, 
B.~Adeva$^{36}$, 
M.~Adinolfi$^{45}$, 
C.~Adrover$^{6}$, 
A.~Affolder$^{51}$, 
Z.~Ajaltouni$^{5}$, 
J.~Albrecht$^{9}$, 
F.~Alessio$^{37}$, 
M.~Alexander$^{50}$
% ...

{\footnotesize \it
% ...
$ ^{5}$Clermont Universit\'{e}, Universit\'{e} Blaise Pascal, CNRS/IN2P3, LPC, Clermont-Ferrand, France\\
$ ^{6}$CPPM, Aix-Marseille Universit\'{e}, CNRS/IN2P3, Marseille, France\\
% ...
$ ^{9}$Fakult\"{a}t Physik, Technische Universit\"{a}t Dortmund, Dortmund, Germany\\
% ...
$ ^{35}$Universitat de Barcelona, Barcelona, Spain\\
$ ^{36}$Universidad de Santiago de Compostela, Santiago de Compostela, Spain\\
$ ^{37}$European Organization for Nuclear Research (CERN), Geneva, Switzerland\\
% ...
$ ^{40}$Nikhef National Institute for Subatomic Physics, Amsterdam, The Netherlands\\
% ...
$ ^{45}$H.H. Wills Physics Laboratory, University of Bristol, Bristol, United Kingdom\\
% ...
$ ^{50}$School of Physics and Astronomy, University of Glasgow, Glasgow, United Kingdom\\
$ ^{51}$Oliver Lodge Laboratory, University of Liverpool, Liverpool, United Kingdom\\
% ...
$ ^{n}$LIFAELS, La Salle, Universitat Ramon Llull, Barcelona, Spain\\
% ...
}
\end{flushleft}

\cleardoublepage


\renewcommand{\thefootnote}{\arabic{footnote}}
\setcounter{footnote}{0}



\pagestyle{plain} 
\setcounter{page}{1}
\pagenumbering{arabic}


%

\section{Introduction}
\label{sec:Introduction}

The combined symmetry of charge conjugation and parity (\CP) is broken in the
weak interaction of the Standard Model by a single phase in the
Cabibbo-Kobayashi-Maskawa matrix~\cite{Cabibbo:1963yz,
  Kobayashi:1973fv}. Physics beyond the Standard Model may reveal itself in the
form of additional sources of \CP violation. In both the \Kz and \Bd systems \CP
violation has been unambiguously observed, and is in agreement with the Standard
Model predictions. In contrast, \CP violation in the charm sector has yet to
be established. The amount of \CP violation in charm decays was generally
expected to be much smaller than the 1\% level in the Standard
Model~\cite{Bianco:2003vb, Grossman:2006jg}. The \lhcb collaboration, however,
reported evidence with $3.5$ standard deviations significance for direct \CP
violation in two-body, singly-Cabibbo-suppressed \Dz
decays~\cite{LHCb-PAPER-2011-023}. The difference in \CP asymmetries between
\dkk and \dpipi decays was found to be
$\DACP=(-0.82\pm0.21\stat\pm0.11\syst)\%$. This result sparked a theoretical
debate on whether or not this could be accommodated within the Standard
Model. For a comprehensive review see Ref.~\cite{LHCb-PAPER-2012-031}.

After the \lhcb paper, the \cdf and \belle collaborations presented
measurements of
$\DACP=(-0.62\pm0.21\stat\pm0.10\syst)\%$~\cite{Collaboration:2012qw} and
$\DACP=(-0.87\pm0.41\stat\pm0.06\syst)\%$~\cite{Ko:2012px}, respectively. These
numbers are included in the average from the Heavy Flavor Averaging Group
(HFAG)~\cite{Amhis:2012bh}, together with a previous
measurement~\cite{Aubert:2007if} from the \babar collaboration, yielding a world
average of the difference in direct \CP violation of $\Delta a_{\CP}^{\rm
dir}=(-0.68\pm0.15)\%$.\footnote{The relation between \DACP and $\Delta
a_{\CP}^{\rm dir}$ is explained in Sect.~\ref{sec:lifetime}.}

In all previous results $\Dstarp\to\Dz\pip$ decays\footnote{The inclusion of
  charge-conjugated modes is implied throughout this paper, unless explicitly
  stated otherwise.} have been used as the source of the \Dz sample, and the
emitted pion was used to determine the flavour of the neutral \PD meson (\ie,
whether it is \Dz or \Dzb). In this paper a measurement of \DACP is presented
using \Dz mesons produced in semileptonic \bquark-hadron decays where the
flavour of the neutral \PD meson is tagged by the accompanying charged
lepton. This approach provides an independent determination of \DACP.

\section{Method and formalism}
\label{sec:method}

The measured (raw) asymmetry for a \Dz decay to a \CP eigenstate $f$
is defined as
\begin{equation}
  \Araw = \frac{N(\Dz\to f)-N(\Dzb\to f)}{N(\Dz\to f)+N(\Dzb\to f)} \ ,
\end{equation}
where $N$ denotes the observed yield for the given decay. The initial flavour of
the neutral \PD meson is tagged by the charge of the accompanying muon in the
semileptonic \bquark-hadron (\PB) decay to the $\PD\mu\PX$ final state. A
positive muon is associated with a \Dzb meson, and a negative muon with a \Dz
meson. The \PX denotes any other particle(s) produced in the semileptonic \PB
decay, which are not reconstructed (\eg, the neutrino).

The raw asymmetry can be written in terms of the \Dz decay rate, $\Gamma$, the
muon detection efficiency, $\varepsilon$, and the \Dz production rate in
semileptonic \bquark-hadron decays, ${\cal P}$, as
\begin{equation}
  \Araw = \frac{\Gamma(\Dz)\varepsilon(\mun){\cal P}(\Dz) - 
    \Gamma(\Dzb)\varepsilon(\mup){\cal P}(\Dzb)}
	{\Gamma(\Dz)\varepsilon(\mun){\cal P}(\Dz) + 
	  \Gamma(\Dzb)\varepsilon(\mup){\cal P}(\Dzb)} \;\;\; .
\end{equation}
Defining the \CP asymmetry as 
$\ACP = (\Gamma(\Dz)-\Gamma(\Dzb))/(\Gamma(\Dz)+\Gamma(\Dzb))$,
the muon detection asymmetry as $\AD=
(\varepsilon(\mun)-\varepsilon(\mup))/(\varepsilon(\mup)+\varepsilon(\mun))$, 
and 
the effective production asymmetry as 
$\AP = ({\cal P}(\Dz)-{\cal P}(\Dzb))/({\cal P}(\Dz)+{\cal P}(\Dzb))$,
the raw asymmetry can be written to first order as
\begin{equation}
  \Araw \approx \ACP + \AD + \AP \ .
  \label{eq:Araw}
\end{equation}
The effective production asymmetry is due to different production rates of
\bquark- and \bquarkbar-hadrons and also includes any effect due to semileptonic
asymmetries in neutral \B mesons. As the detection and production asymmetries
are of order 1\%, the approximation in Eq.~(\ref{eq:Araw}) is valid up to
corrections of order $10^{-6}$. Both detection and production asymmetries differ
from those in the analyses using \Dstarpm decays, where the \Dstarpm mesons are
produced directly in the primary $pp$ interaction. In these ``prompt'' decays a
possible detection asymmetry enters through the reconstruction of the tagging
pion, and the production asymmetry is that of the prompt \Dstarpm mesons.

By taking the difference between the raw asymmetries measured in the \dkk and
\dpipi decays the detection and production asymmetries cancel, giving a
robust measurement of the \CP asymmetry difference
\begin{equation}
  \DACPraw =  \AKK - \Apipi \approx \AcpKK - \Acppipi \ .
  \label{eq:delta}
\end{equation}
Since the detection and the production depend on the kinematics of the process
under study, the cancellation is only complete when the kinematic distributions
of the muon and \bquark-hadron are the same for both \dkk and \dpipi. A
weighting procedure is used to improve the cancellation by equalising the
kinematic distributions.

\section{Detector and simulation}
\label{sec:Detector}

The \lhcb detector~\cite{Alves:2008zz} is a single-arm forward spectrometer
covering the \mbox{pseudorapidity} range $2<\eta <5$, designed for the study of
particles containing \bquark or \cquark quarks. The detector includes a
high-precision tracking system consisting of a silicon-strip vertex detector
surrounding the $pp$ interaction region, a large-area silicon-strip detector
located upstream of a dipole magnet with a bending power of about $4{\rm\,Tm}$,
and three stations of silicon-strip detectors and straw drift tubes placed
downstream. The polarity of the magnet is reversed repeatedly during data
taking, which causes all detection asymmetries that are induced by the
left--right separation of charged particles to change sign. The combined
tracking system has momentum resolution $\Delta p/p$ that varies from 0.4\% at
5\gevc to 0.6\% at 100\gevc, and impact parameter resolution of 20\mum for
tracks with high transverse momentum. Charged hadrons are identified using two
ring-imaging Cherenkov detectors~\cite{LHCb-DP-2012-003}. Muons are identified
by a system composed of alternating layers of iron and multiwire proportional
chambers. The trigger~\cite{LHCb-DP-2012-004} consists of a hardware stage,
based on information from the calorimeter and muon systems, followed by a
software stage, which applies a full event reconstruction.

In the simulation, $pp$ collisions are generated using
\pythia~6.4~\cite{Sjostrand:2006za} with a specific \lhcb
configuration~\cite{LHCb-PROC-2010-056}. Decays of hadronic particles are
described by \evtgen~\cite{Lange:2001uf} in which final state radiation is
generated using \photos~\cite{Golonka:2005pn}. The interaction of the generated
particles with the detector and its response are implemented using the \geant
toolkit~\cite{Allison:2006ve, *Agostinelli:2002hh} as described in
Ref.~\cite{LHCb-PROC-2011-006}. The \Bp and \Bz mesons in the simulated events
are forced to decay semileptonically using a cocktail of decay modes, including
those that involve excited \PD states and \Ptau leptons, that lead to final
states with a \Dz meson and a muon.

\section{Data set and selection}

This analysis uses the \lhcb 2011 data set, corresponding to an integrated
luminosity of $1.0\invfb$, of which $0.4\invfb$ is taken with the magnet field
pointing up and $0.6\invfb$ with the magnet field pointing down. The measurement
of \DACP is performed separately for the two field polarities. The final value
for \DACP is obtained by taking the arithmetic mean of the two results to reduce
as much as possible any residual effect of the detection asymmetry. To minimise
potential trigger biases the candidates are required to be accepted by specific
trigger decisions. About $87\%$ of the candidates in the final selection are
triggered at the hardware stage by the muon system only, about $3\%$ by the
hadronic calorimeter only and about $10\%$ by both. The muon trigger requires
the muon transverse momentum, \pt, to be greater than $1.48\gevc$. The effect of
a charge-dependent shift in the \pt estimate in this trigger is corrected, which
requires tightening the muon transverse momentum cut, as measured by the
hardware trigger, to $\pt>1.64\gevc$.  In the software trigger the candidates
are selected by either a single muon trigger or by a topological trigger, which
selects combinations of a muon with one or two additional tracks that are
consistent with the topological signature of \bquark-hadron decays. At this
level, $5\%$ of the candidates in the final selection are selected by the single
muon trigger only, $79\%$ by the topological trigger only, and $16\%$ by both.

In order to suppress backgrounds, the $\chi^2$ per degree of freedom of the
track fit is required to be smaller than 4 for the kaons and pions and smaller
than 5 for the muon. Furthermore, the $\chi^2$ per degree of freedom of each of
the \bquark-hadron and \Dz decay vertex fits is required to be smaller than 6,
and the impact parameter $\chi^2$ (defined as the difference between the
$\chi^2$ of the primary vertex formed with and without the considered tracks) is
required to be larger than 9 for all three tracks. The significance of the
distance between the primary vertex and the \Dz decay vertex is required to be
above 10. The momentum and transverse momentum of the muon are required to be
above $3\gevc$ and $1.2\gevc$,\footnote{This cut affects mainly the candidates
  triggered by the hadronic calorimeter at the hardware level.} and the momentum
and transverse momentum of the \Dz daughters above $2\gevc$ and $0.3\gevc$. The
\Dz transverse momentum must be above $0.5\gevc$ and the scalar \pt sum of its
daughters above $1.4\gevc$. The invariant mass of the \Dz--muon combination is
required to be between $2.5$ and $5.0\gevcc$ to suppress background. The upper
bound removes three-body final state \bquark-hadron decays. The reconstructed
decay time of the \Dz meson (measured from the \bquark-hadron decay vertex) is
required to be positive. The requirement on the muon impact parameter reduces
the contribution from \Dz mesons produced directly in the $pp$ collision to
below $3\%$. Requirements on the \Dz decay topology are minimal in order to keep
the lifetime acceptance similar for the \dkk and \dpipi modes.

A potentially significant background from $\B\to\jpsi X$ decays is suppressed by
removing candidates where the invariant mass of the muon and the
oppositely-charged \Dz daughter is within three times the mass resolution from
the \jpsi or \psitwos mass and the \Dz daughter passes muon identification
requirements. Reflections from Cabibbo-favoured \dkpi decays are observed in the
mass regions below and above the signal peaks in the \dpipi and \dkk samples,
respectively. Information from the relevant detectors in LHCb is combined into
differences between the logarithms of the particle identification likelihoods
under different mass hypotheses (DLL). The selected kaons are required to have
$\dllkpi\equiv\ln{\cal L}_K-\ln{\cal L}_\pi>10$ and the selected pions are
required to have $\dllkpi<-2$. The \dkpi mode is used as a control channel and
is selected with the same requirements as the two decay modes of interest.

\section{Determination of the asymmetries}

\begin{figure}
  \begin{center}
    \includegraphics[width=0.49\textwidth]{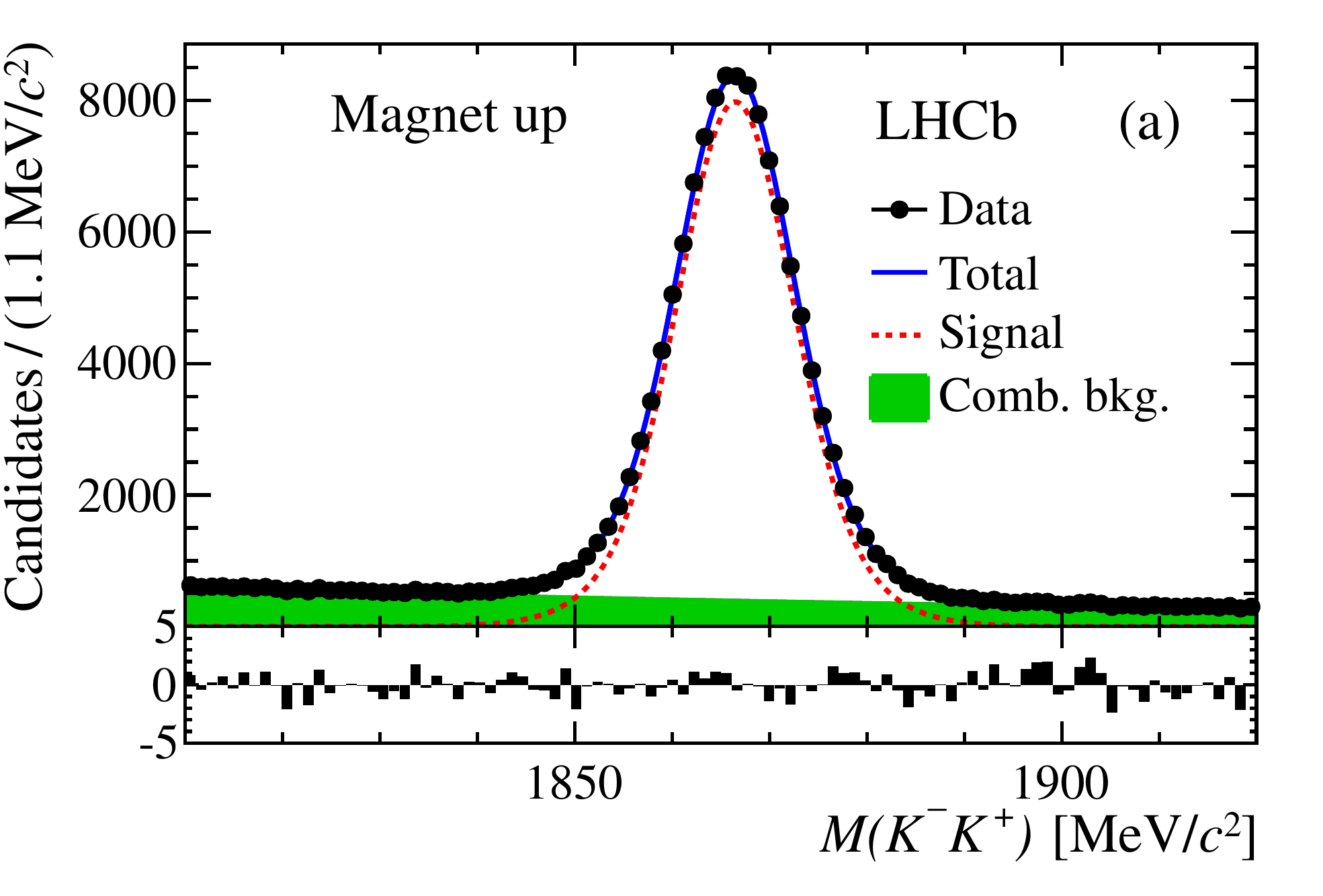}
    \includegraphics[width=0.49\textwidth]{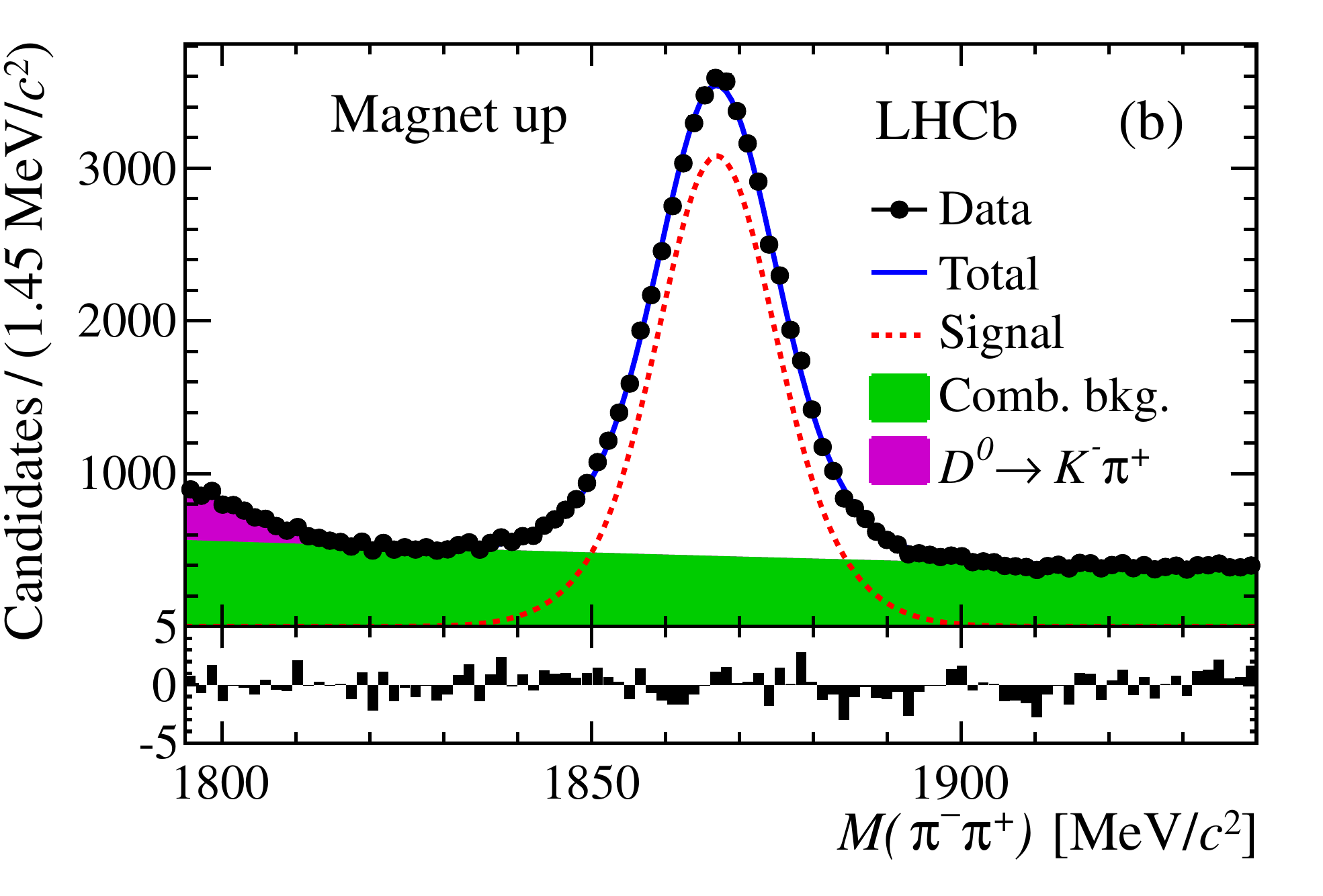}
    \includegraphics[width=0.49\textwidth]{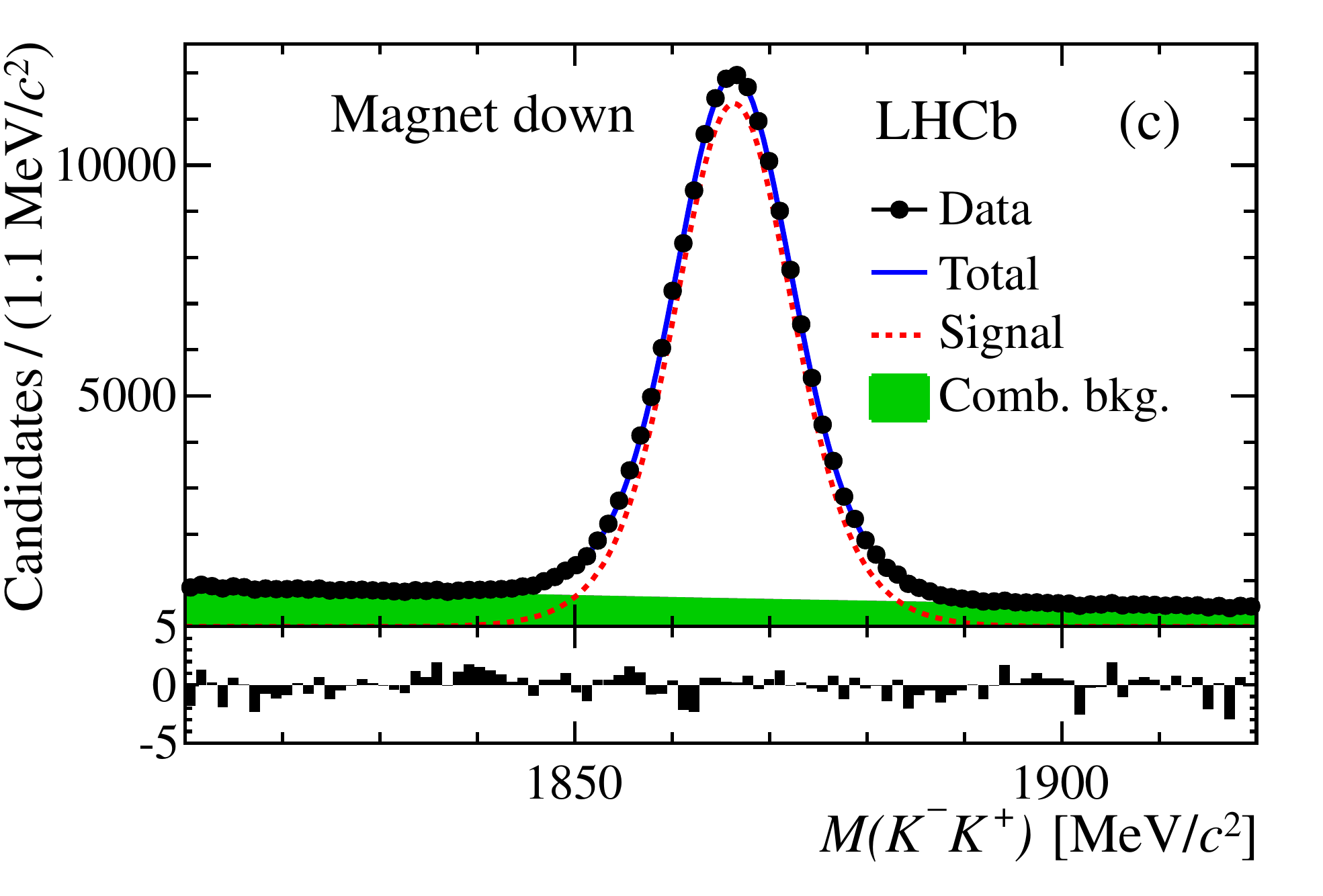}
    \includegraphics[width=0.49\textwidth]{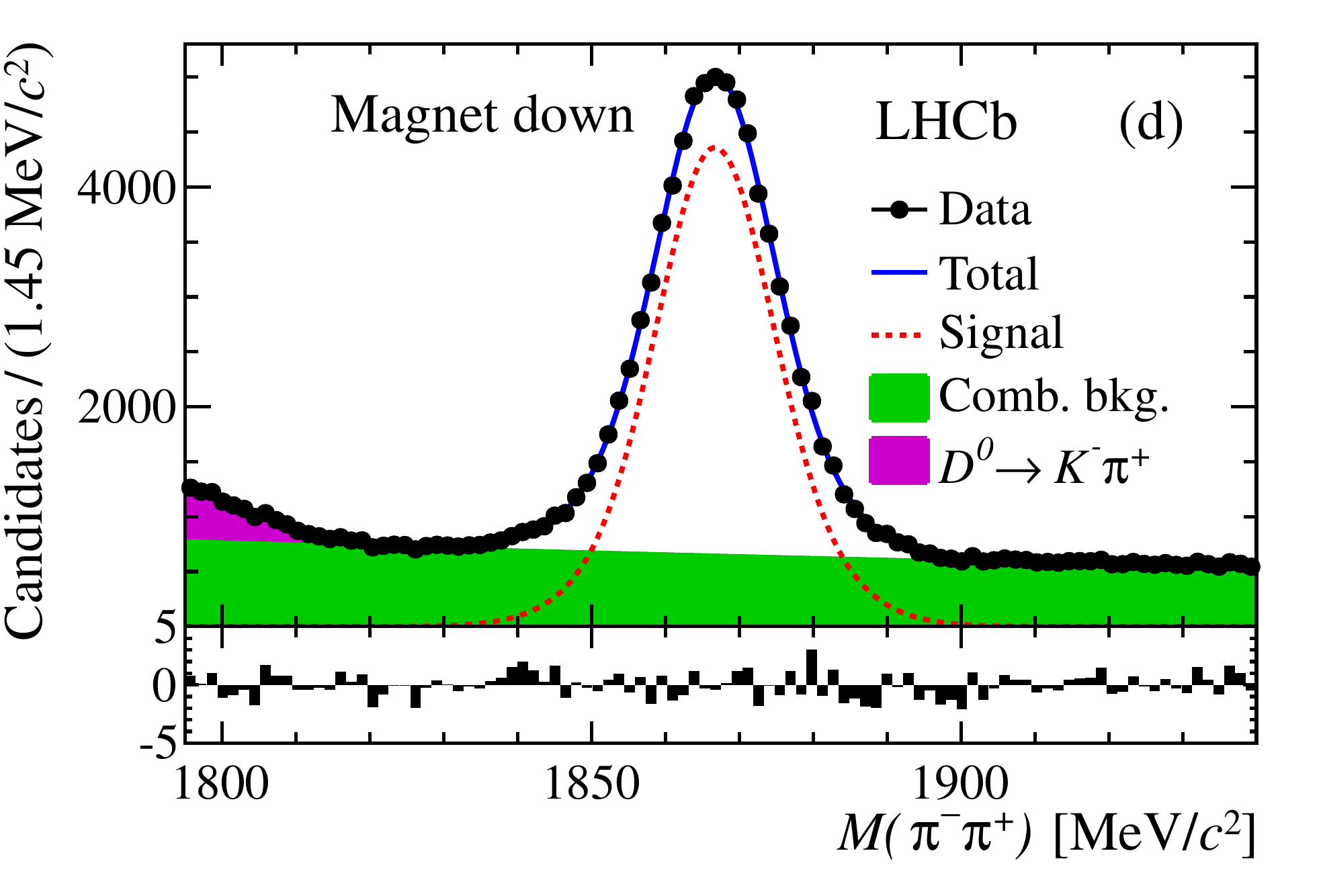}
  \end{center}
  \vspace*{-1.0cm}
  \caption{\small Invariant mass distributions for (a, c) \dkk and (b, d) \dpipi
    muon-tagged candidates for the two magnet polarities. The result of the
    fit is overlaid, showing the contribution from signal,
    combinatorial background and \dkpi reflection. Underneath each plot the pull
    in each mass bin is shown.}
  \label{fig:default_asymmetry}
\end{figure}

The invariant mass distributions for the muon-tagged \Dz candidates are shown in
Fig.~\ref{fig:default_asymmetry}. To determine the numbers of signal candidates
after selection, a binned maximum likelihood fit to each of these distributions
is performed. The signal is modelled by the sum of two Gaussian functions with
common means, but different widths. The combinatorial background is described by
an exponential shape. For the $\pim\pip$ invariant mass distribution the fit is
performed in the range between $1795-1940\mevcc$ and a Gaussian distribution is
used to model the tail of the reflection from \dkpi decays. For the $\Km\Kp$
invariant mass distribution the fit range is restricted to $1810-1920\mevcc$
such that the contamination from the \dkpi reflection and from partially
reconstructed $\dkk\piz$ and $\Dp\to\Km\Kp\pip$ decays is negligible. The total
number of signal candidates determined from the fit is $(558.9\pm0.9)\times10^3$
for \dkk decays and $(221.6\pm0.8)\times10^3$ for \dpipi decays.

The raw asymmetries are determined with simultaneous binned likelihood fits to
the \Dz mass distributions for positive and negative muon tags where the shape
parameters for the signal and the \dkpi reflection are required to be the
same. The background shape can vary independently for positive and negative muon
tags. Table~\ref{tab:default_asymmetry} lists the raw asymmetries for both
modes, and for the \dkpi control mode. An additional asymmetry in the \dkpi mode
originates from the different cross-sections in matter for positive and negative
kaons. It can be seen that the asymmetry in this mode is consistent for the two
magnetic field polarities, which indicates that the detection asymmetry related
to the magnetic field is at most $\order(10^{-3})$.

\begin{table}[h]
  \begin{center}
    \caption{\small Unweighted raw asymmetries (in \%) for the \dpipi,
      \dkk and \dkpi decays for the two magnet polarities. The mean value is the
      arithmetic average over the two polarities. The uncertainties are
      statistical only.}
  \label{tab:default_asymmetry}
  \vspace{0.1cm}
  \begin{tabular}{l|r@{$\pm$}lr@{$\pm$}lr@{$\pm$}l} &
\multicolumn{2}{c}{Magnet up} & \multicolumn{2}{c}{Magnet down} & \multicolumn{2}{c}{Mean} \\
\hline
    \AKKUnw   & $-0.33$&$0.23$ & $-0.22$&$0.19$ & $-0.28$&$0.15$ \\
    \ApipiUnw & $-1.18$&$0.40$ & $-0.35$&$0.34$ & $-0.77$&$0.26$ \\
    \hline
    \DACPUnw  &  $0.85$&$0.46$ &  $0.13$&$0.39$ &  $0.49$&$0.30$ \\
    \hline
    \AKpiUnw  & $-1.64$&$0.10$ & $-1.60$&$0.08 $ & $-1.62$&$0.06$\\
  \end{tabular} \end{center}
\end{table}

\subsection{Differences in kinematic distributions}
\label{sec:weighting}

\begin{figure}
  \begin{center}
    \includegraphics[width=0.49\textwidth]{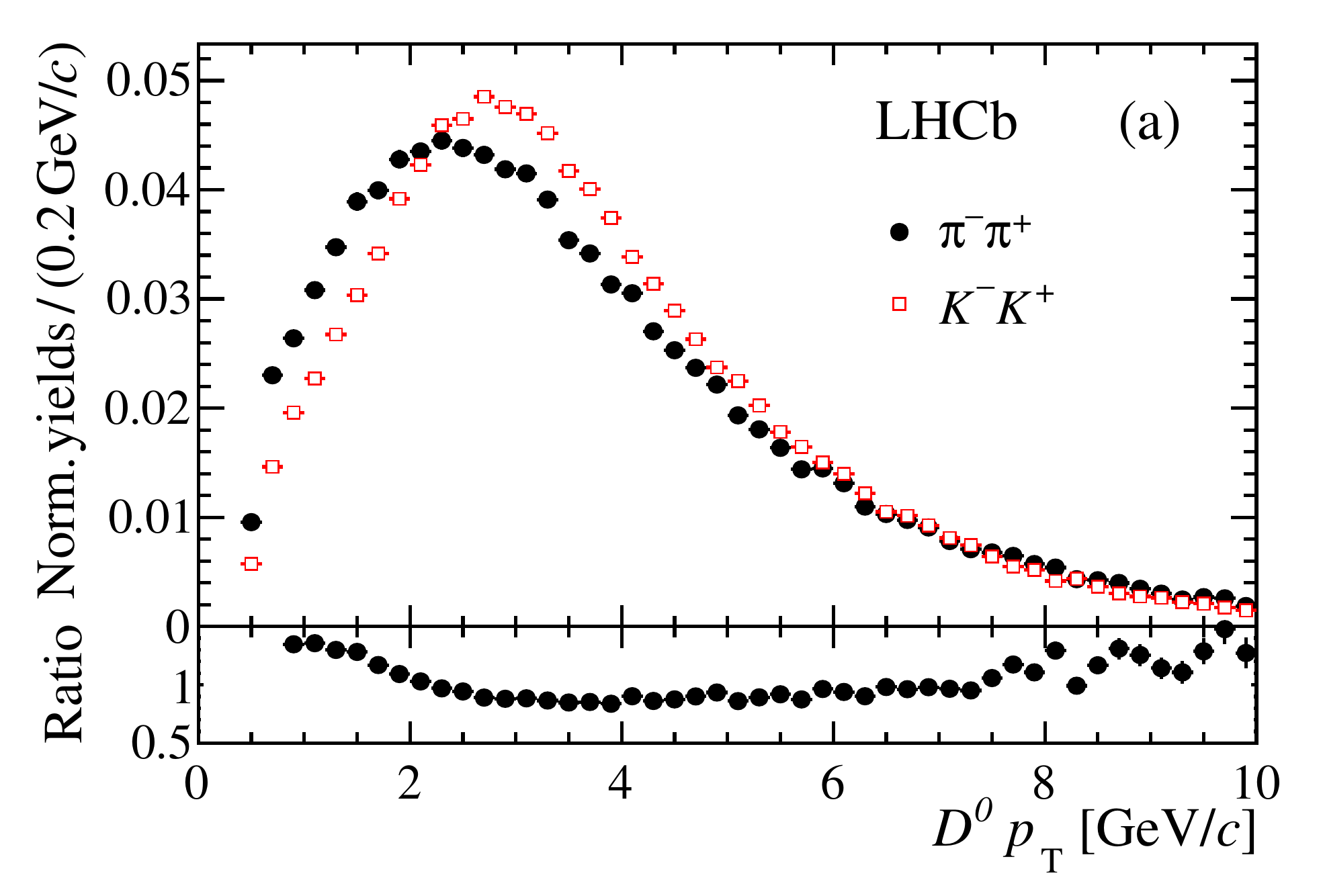}
    \includegraphics[width=0.49\textwidth]{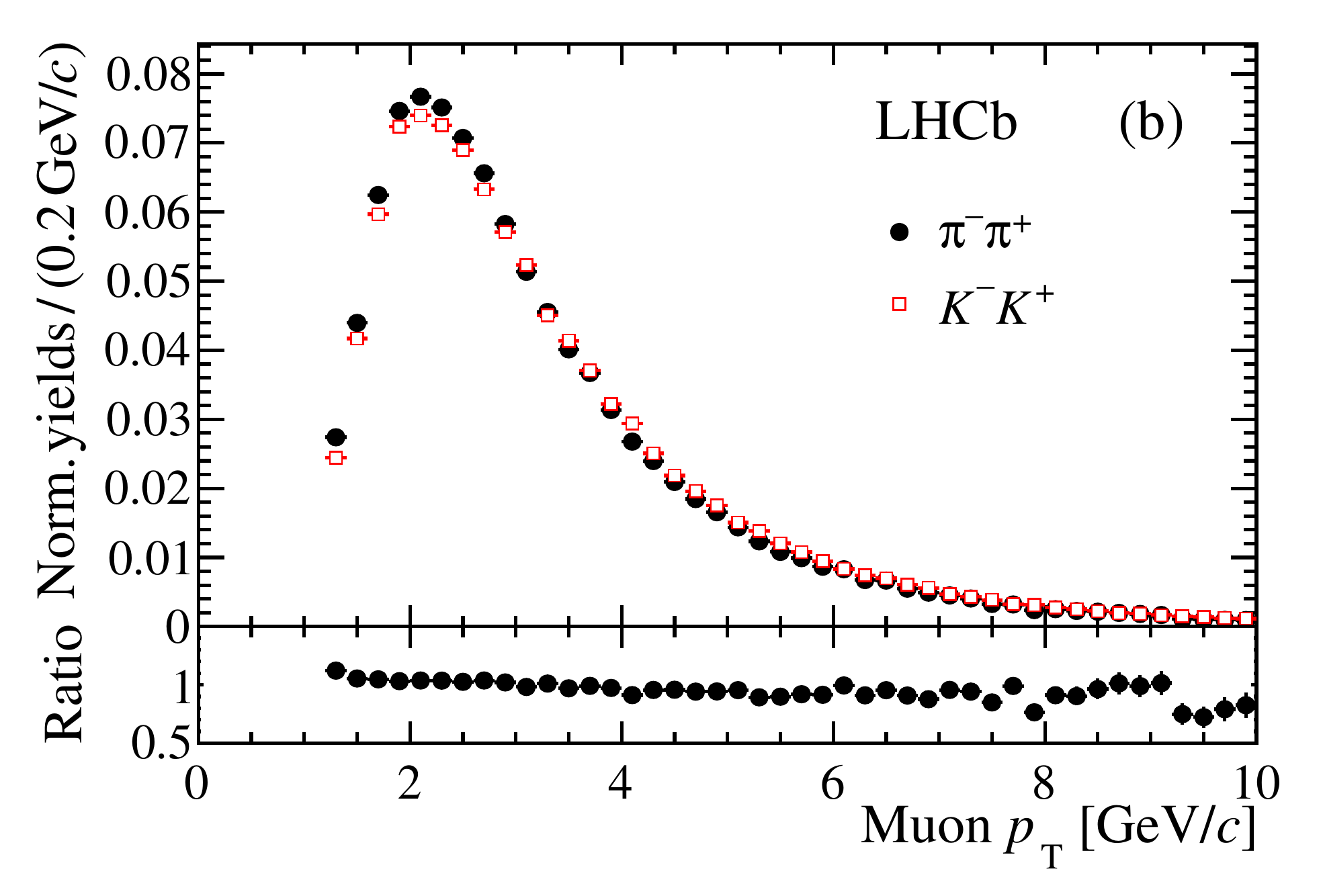}
    \includegraphics[width=0.49\textwidth]{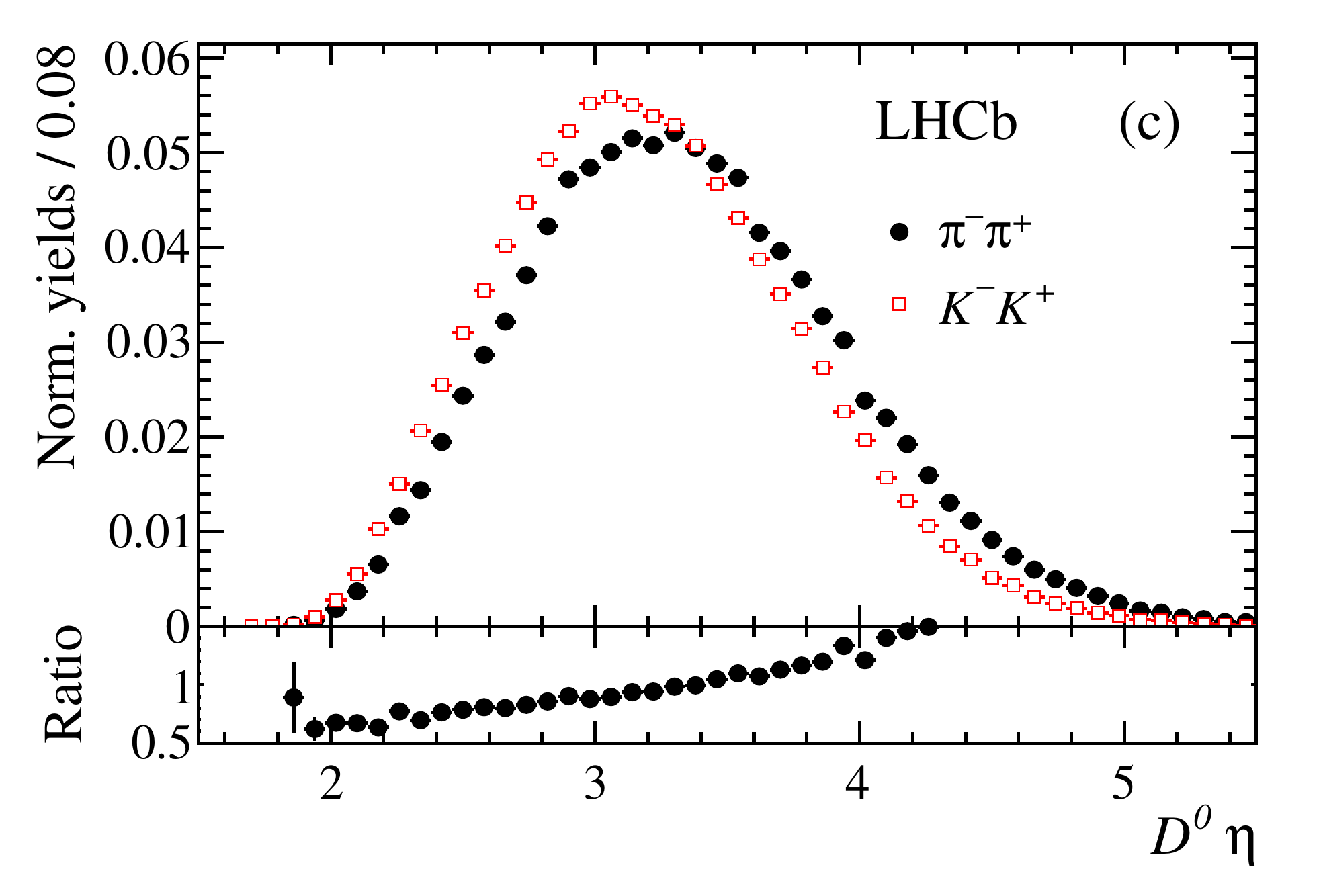}
    \includegraphics[width=0.49\textwidth]{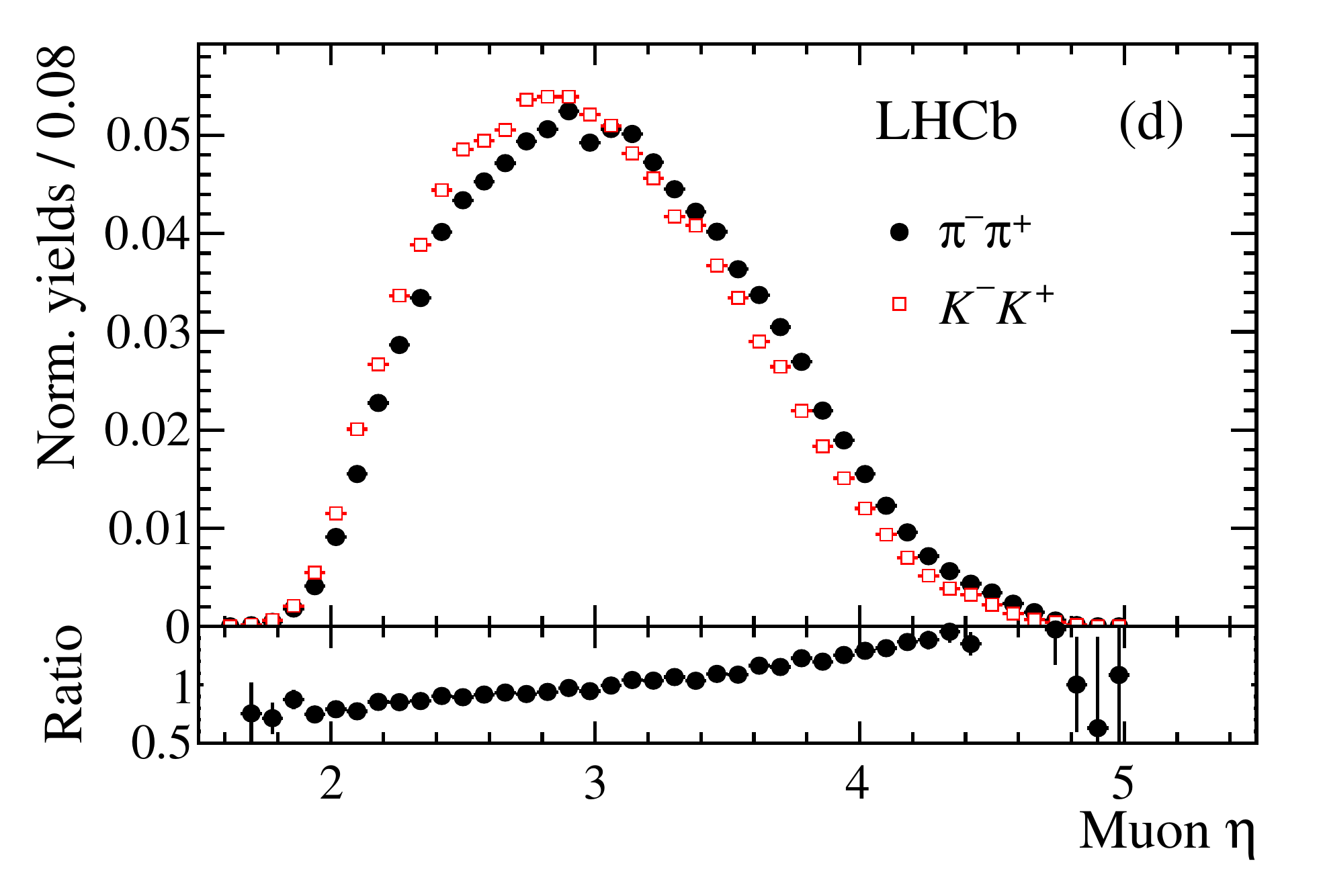}
    \end{center}
    \vspace*{-1.0cm}
    \caption{\small Kinematic distributions of the (a, c) \Dz meson and (b, d)
      muon for \dpipi (black circles) and \dkk (red squares) candidates
      normalised to unit area. The histograms show the distributions of signal
      candidates, after background subtraction. Underneath each plot the ratio
      of the two distributions is shown.}
    \label{fig:kinematics_down}
\end{figure}

\begin{figure}
  \begin{center}
    \includegraphics[width=0.49\textwidth]{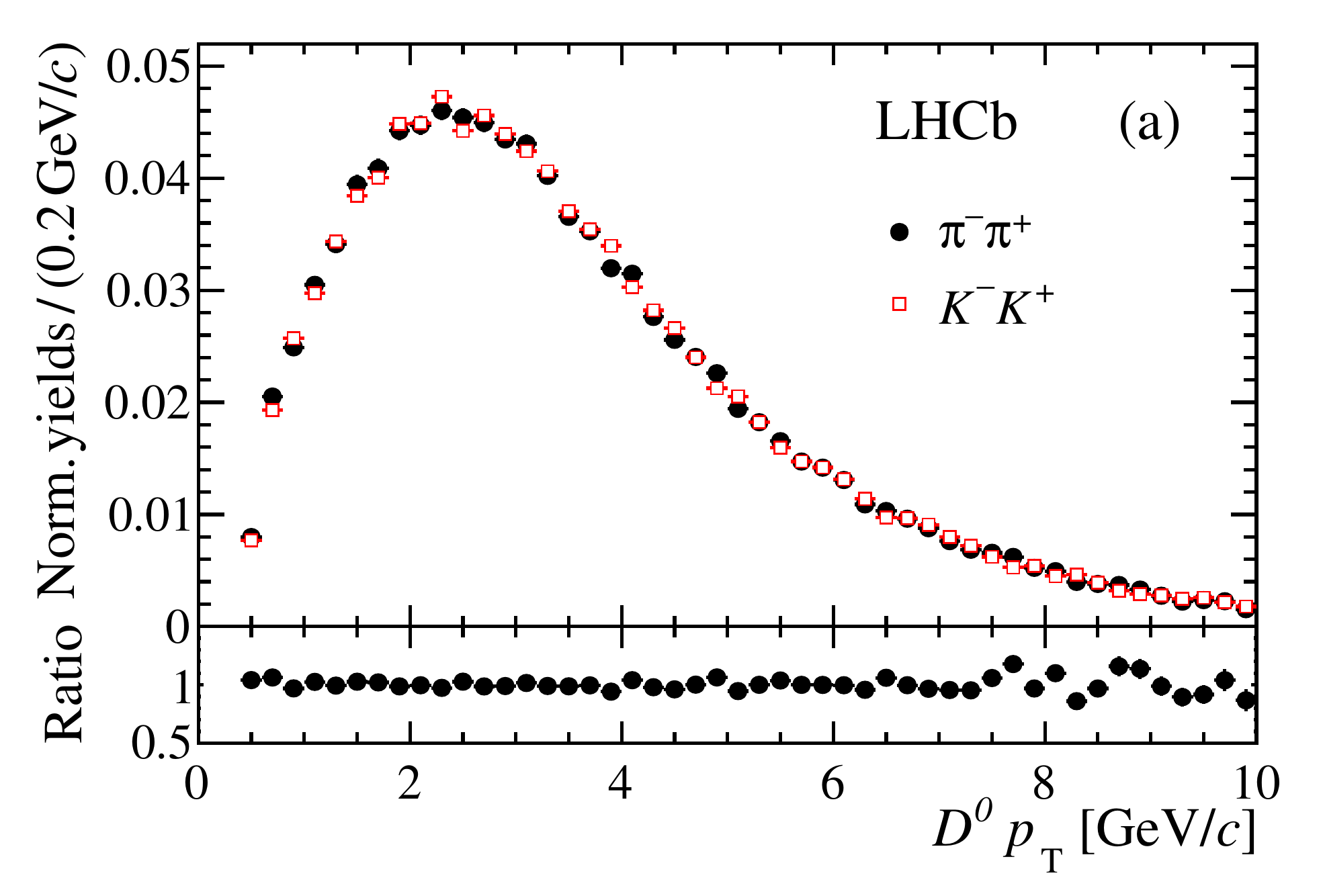}
    \includegraphics[width=0.49\textwidth]{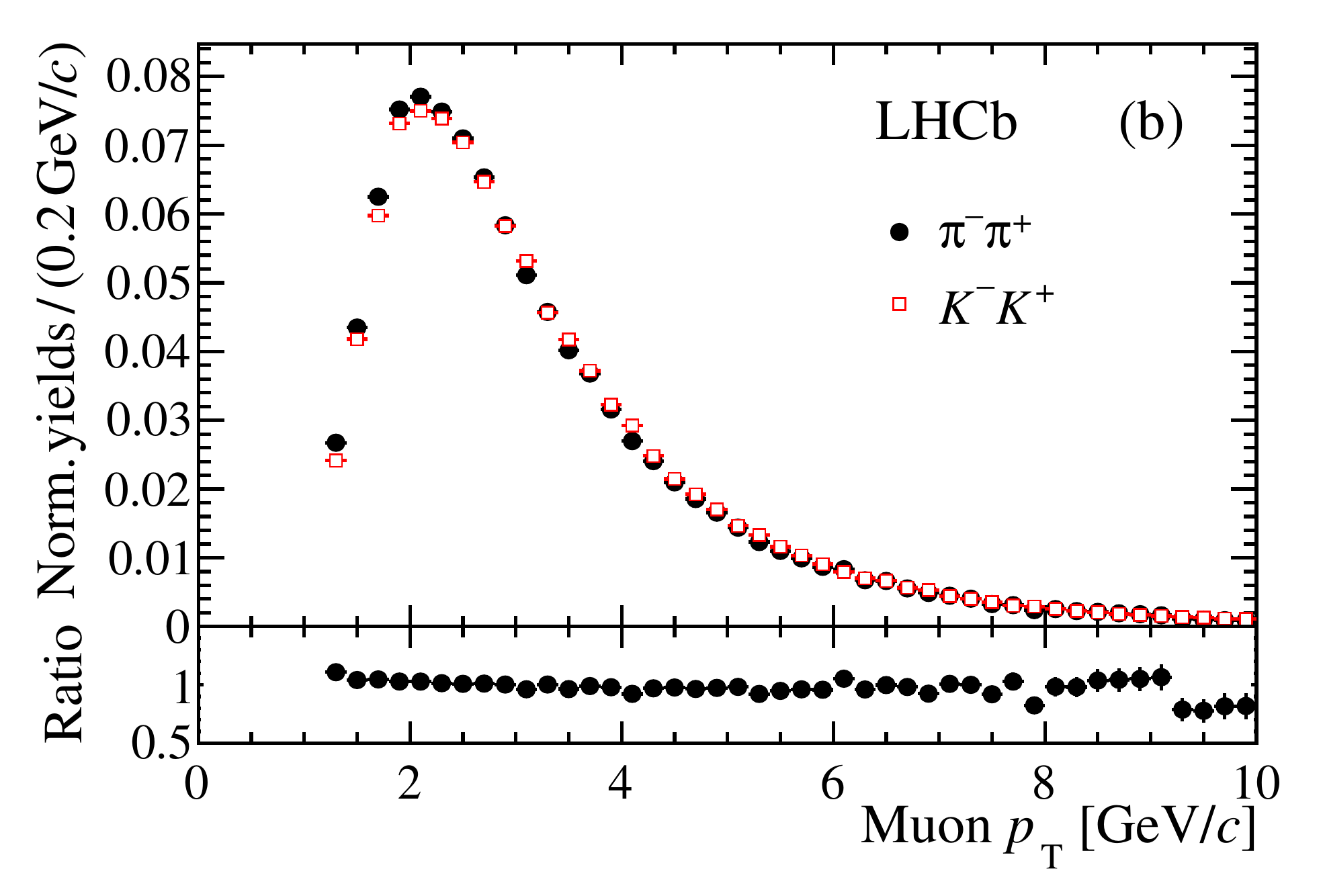}
    \includegraphics[width=0.49\textwidth]{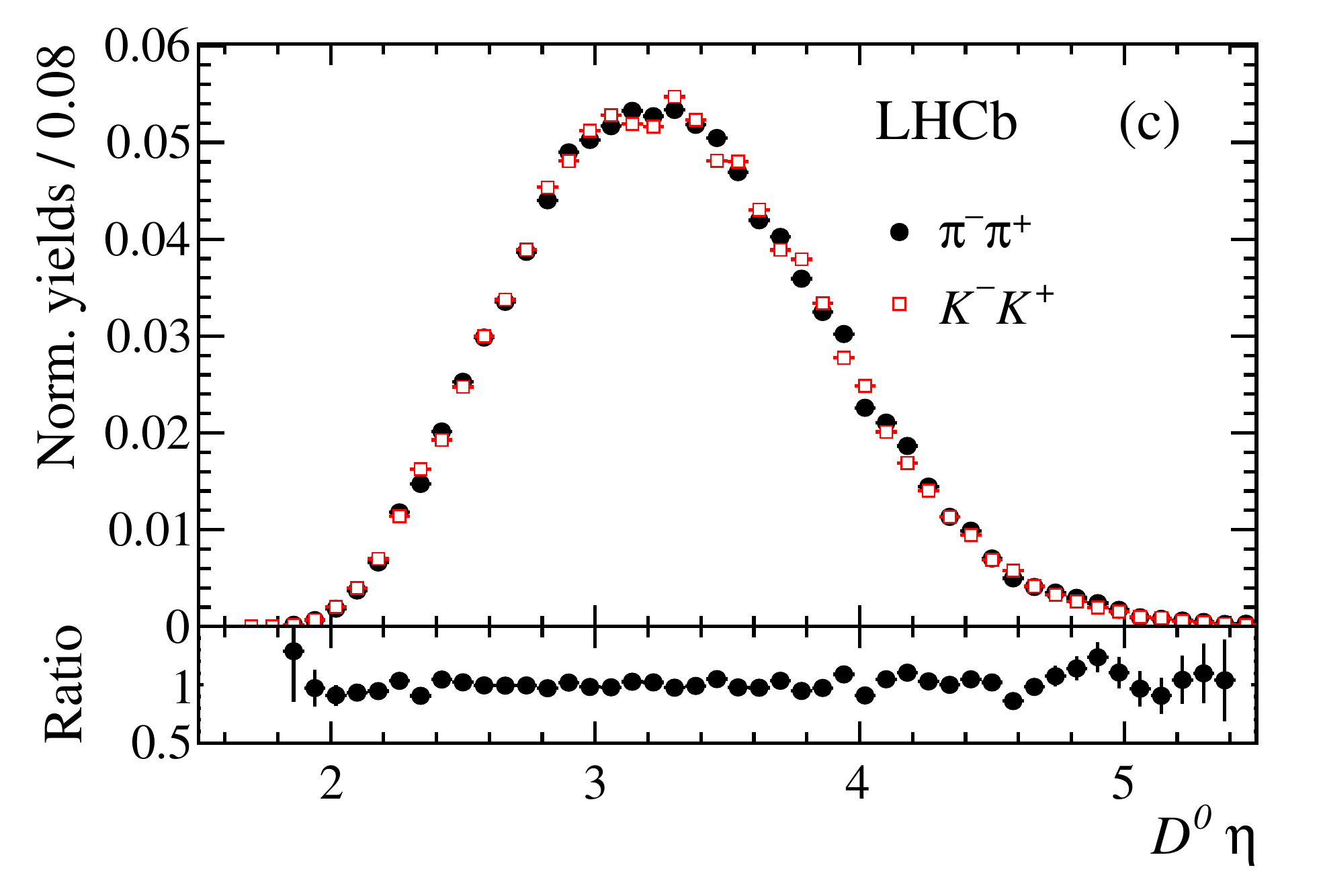}
    \includegraphics[width=0.49\textwidth]{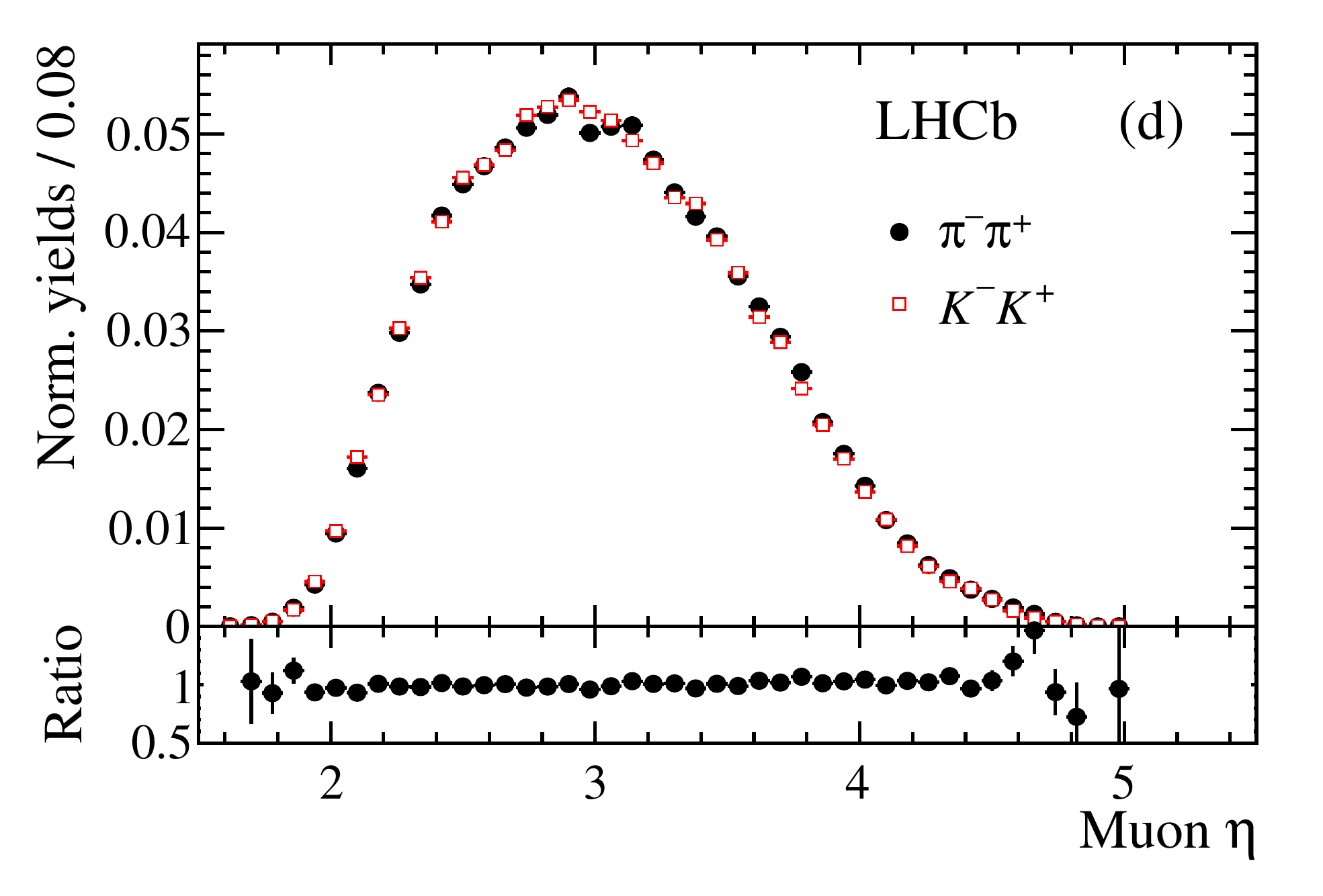}
    \end{center}
    \vspace*{-1.0cm}
    \caption{\small Kinematic distributions of the (a, c) \Dz meson and (b, d)
      muon for \dpipi (black circles) and \dkk (red squares) candidates
      normalised to unit area after the weighting procedure. The histograms show
      the distributions of signal candidates, after background
      subtraction. Underneath each plot the ratio of the two distributions is
      shown.}
    \label{fig:kinematics_weighted_down}
\end{figure}

Since the detection and production asymmetries may have kinematic dependences,
the cancellation in Eq.~(\ref{eq:delta}) is only valid if the kinematic
distributions of the muon and \bquark-hadron are similar for both \dkk and
\dpipi decays. After the trigger and selection requirements the kinematic
distributions for the two decay modes are, however, slightly different. Although
the selection is largely the same, the particle identification requirements
introduce differences in the momentum distributions. In addition, due to the
difference in available phase space, the pions in \dpipi decays have a harder
momentum spectrum compared to the kaons in \dkk decays. The muon trigger and
selection requirements are identical. Nevertheless, the \Dz meson and the muon
are kinematically correlated since they originate from the same decay, causing
also the muon kinematic distributions to be different for the two decay
modes. Figure~\ref{fig:kinematics_down} shows the \pt and pseudorapidity $\eta$
distributions for the \Dz meson and the muon.  The background has been
statistically subtracted using the \sPlot~method~\cite{Pivk:2004ty}. In order to
obtain the same kinematic distributions for both decays, the \Dz candidates are
given a weight depending on their \pt and $\eta$ values.  The weights are
obtained from a comparison of the background-subtracted distributions and are
applied to either \dkk or \dpipi candidates depending on which has most events
in the given kinematic bin.  Figure~\ref{fig:kinematics_weighted_down} shows the
weighted kinematic distributions for both decay modes. Whereas the weights are
determined purely on the basis of the \Dz \pt and $\eta$ distributions, after
the weighting, the muon distributions are also in excellent agreement. The raw
asymmetries after the weighting procedure for the \dkk and \dpipi modes are
given in Table~\ref{tab:default_asymmetry_weighted}. There are minor changes in
the values of the raw asymmetries and \DACP with respect to the unweighted
results, showing that the effect of the difference in kinematic distributions is
small.

\begin{table}[h]
  \begin{center}
    \caption{\small Weighted raw asymmetries (in \%) for the \dpipi and
      \dkk decays for the two magnet polarities. The mean value is
      the arithmetic average over the two polarities. The uncertainties are
      statistical only.}
    \label{tab:default_asymmetry_weighted}
  \vspace{0.1cm}
  \begin{tabular}{l|r@{$\pm$}lr@{$\pm$}lr@{$\pm$}l} &
\multicolumn{2}{c}{Magnet up} & \multicolumn{2}{c}{Magnet down} & \multicolumn{2}{c}{Mean} \\
\hline
    \AKK     & $-0.39$&$0.23$ & $-0.20$&$0.19$ & $-0.29$&$0.15$ \\
    \Apipi   & $-1.25$&$0.40$ & $-0.29$&$0.34$ & $-0.77$&$0.26$ \\
    \hline
    \DACPraw &  $0.86$&$0.46$ &  $0.09$&$0.39$ &  $0.48$&$0.30$ \\
  \end{tabular} \end{center}
\end{table}

\subsection{Wrong flavour tags}
\label{sec:mistag}

In some cases the \Dz flavour is not tagged correctly by the muon charge due to
misreconstruction (\eg, a prompt \Dz decay can be combined with a random
muon). The probability to tag a \Dz meson with a positive muon is denoted by
$\mistagP$ and the probability to tag a \Dzb meson with a negative muon by
$\mistagM$. The average mistag probability is $\overline{\mistag}=(\mistagP +
\mistagM)/2$ and the mistag difference is $\Delta\mistag=\mistagP-\mistagM$. The
raw asymmetry in Eq.~(\ref{eq:Araw}) is then modified to
\begin{equation}
  \Araw \approx (1-2\overline{\mistag})(\ACP + \AD + \AP) - \Delta\mistag \ ,
\label{eq:AmeasFinal}
\end{equation}
which makes clear that the average mistag probability dilutes the observed
asymmetry, while any difference in the mistag probability for \Dz and \Dzb gives
rise to a systematic shift in \Araw. Assuming that the values of
$\overline{\mistag}$ and $\Delta\mistag$ are the same for \dkk and \dpipi, the
value of \DACP is then corrected as
\begin{equation}
  \DACP =  (1-2\overline{\mistag})^{-1}(\AKK - \Apipi) \ .
  \label{eq:mistag}
\end{equation}

The mistag probability is estimated from the \dkpi sample. As the \dkpi decay is
almost self-tagging the mistag probability is determined using the charge of the
final state (either $\Kp\pim$ or $\Km\pip$). The wrongly tagged decays include a
fraction of doubly-Cabibbo-suppressed \decay{\Dz}{\Kp\pim} and mixed
$\Dz\to\decay{\Dzb}{\Kp\pim}$ decays. This fraction is calculated to be
$(0.393\pm0.007)\%$ using input from Ref.~\cite{LHCb-PAPER-2012-038}. After
correcting for this fraction the average mistag probability,
$\overline{\mistag}$, is found to be $(0.982\pm0.012)\%$, which means that the
effect from wrong tags constitutes only a small correction on the observed
asymmetries. This number also provides an upper bound of about $2\%$ from any
background from real \Dz decays with a random muon, which includes promptly
produced \Dz decays. The difference in mistag probabilities for \Dz and \Dzb
mesons is found to be $\Delta\mistag=(0.006\pm0.021)\%$ and is neglected.

As a cross-check the mistag probabilities are also determined from a
doubly-tagged sample by reconstructing $\PB\to\Dstarp\mun\PX$ decays where
the \Dstarp decays to $\Dz\pip$ and comparing the
charge of the pion with that of the muon. The fraction of wrongly tagged
decays is estimated from a simultaneous fit, similar to that in
Ref.~\cite{LHCb-PAPER-2011-032}, to the distribution of $\Delta M =
M(h^-h^+\pip)-M(h^-h^+)$ for the full sample and for the wrongly tagged
decays. The mistag probability in the \dkpi sample is $(0.880\pm0.043)\%$, while
the average mistag probability in the \dkk and \dpipi samples equals
$(1.00\pm0.09)\%$. The largest difference with the result obtained from the full
\dkpi sample (\ie, $0.102\%$) is assigned as a systematic uncertainty in the
mistag probability.  The difference in mistag probabilities, $\Delta\mistag$, in
this cross-check is also consistent with zero.

After the weighting and correcting for the mistag probability of
$(0.982\pm0.012\stat\pm0.102\syst)\%$, the difference of the raw asymmetries
between the two modes is found to be
\begin{equation}
  \DACP = (0.49\pm 0.30)\% \ , \nonumber
\end{equation}
where the uncertainty is statistical only. The corresponding systematic
uncertainties are discussed in Sect.~\ref{sec:systematics}.

\section{Measurement of the average decay times}
\label{sec:lifetime}

The time-integrated asymmetry for a decay to a \CP eigenstate $f$ is
defined as
\begin{equation}
  \ACP = \frac{\Gamma(\Dz\to f)-\Gamma(\Dzb\to f)}{\Gamma(\Dz\to f)+
    \Gamma(\Dzb\to f)} \ ,
\end{equation}
where $\Gamma$ is the decay rate for the given channel. As the reconstruction
and selection requirements for the two decay modes are not identical, the decay
time acceptance can be different. This introduces a difference in the
contribution from direct and indirect \CP violation for the two modes. When
assuming the \CP violating phase in \Dz oscillations, $\phi$, to be
universal~\cite{Grossman:2006jg}, the difference between the asymmetries for
\dkk and \dpipi can be written in terms of direct and indirect \CP violation
as~\cite{Gersabeck:2011xj}
\begin{equation}
  \DACP \approx \Delta a_{\CP}^{\rm dir}
  \left(1+y\frac{\overline{\mean{t}}}{\tau}\cos\phi\right) +
  \left(a_{\CP}^{\rm ind}+\overline{a_{\CP}^{\rm dir}} y \cos\phi\right)
  \frac{\Delta\mean{t}}{\tau} \ .
\label{eq:AcpDirectIndirect}
\end{equation}
In this equation the indirect \CP violation is $a_{\CP}^{\rm ind} =
-(A_m/2)y\cos\phi+x\sin\phi$, $x$ and $y$ are the \Dz mixing parameters, $A_m$
represents the \CP violation from mixing, $\tau$ is the average \Dz lifetime,
$\Delta a_{\CP}^{\rm dir}$ and $\overline{a_{\CP}^{\rm dir}}$ are the direct \CP
violation difference and average of the two decay modes, and $\Delta\mean{t}$
and $\overline{\mean{t}}$ are the difference and average of the two mean decay
times. Under SU(3) flavour symmetry, the direct asymmetries in the individual
modes are expected to have opposite sign and therefore add constructively in the
difference. Furthermore, since $y$ is of order $1\%$, $\overline{\mean{t}}/\tau$
is $\order(1)$ and $\Delta\mean{t}/\tau$ is close to zero, \DACP is essentially
equal to the difference in direct \CP violation, $\Delta a_{\CP}^{\rm
  dir}$. While $y$ and $\cos\phi$ can be obtained from the HFAG
averages~\cite{Amhis:2012bh}, in order to interpret \DACP in terms of direct and
indirect \CP violation, the mean decay time $\mean{t}$ in each channel needs to
be measured.

\begin{figure}
  \begin{center}
    \includegraphics[width=0.49\textwidth]{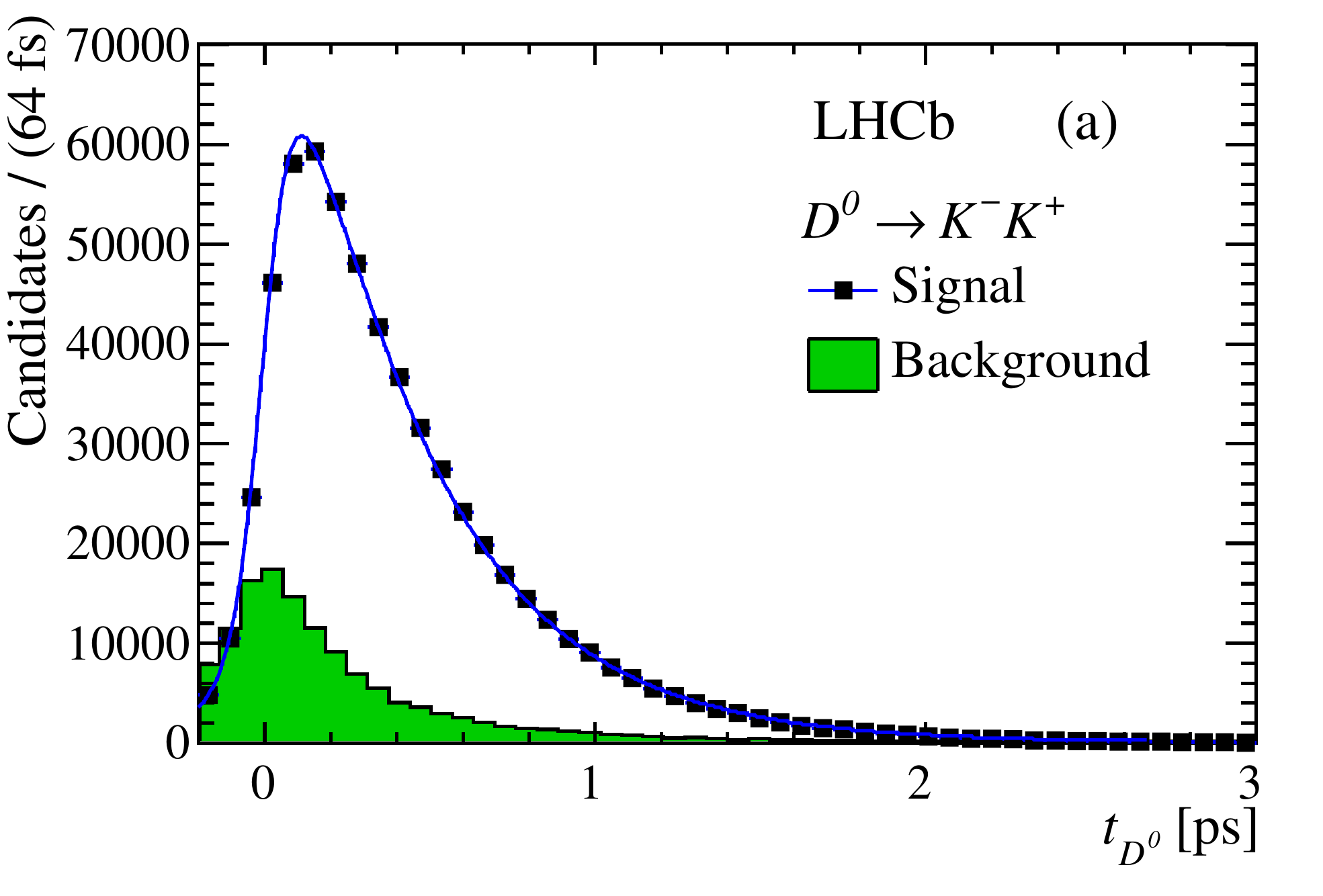}
    \includegraphics[width=0.49\textwidth]{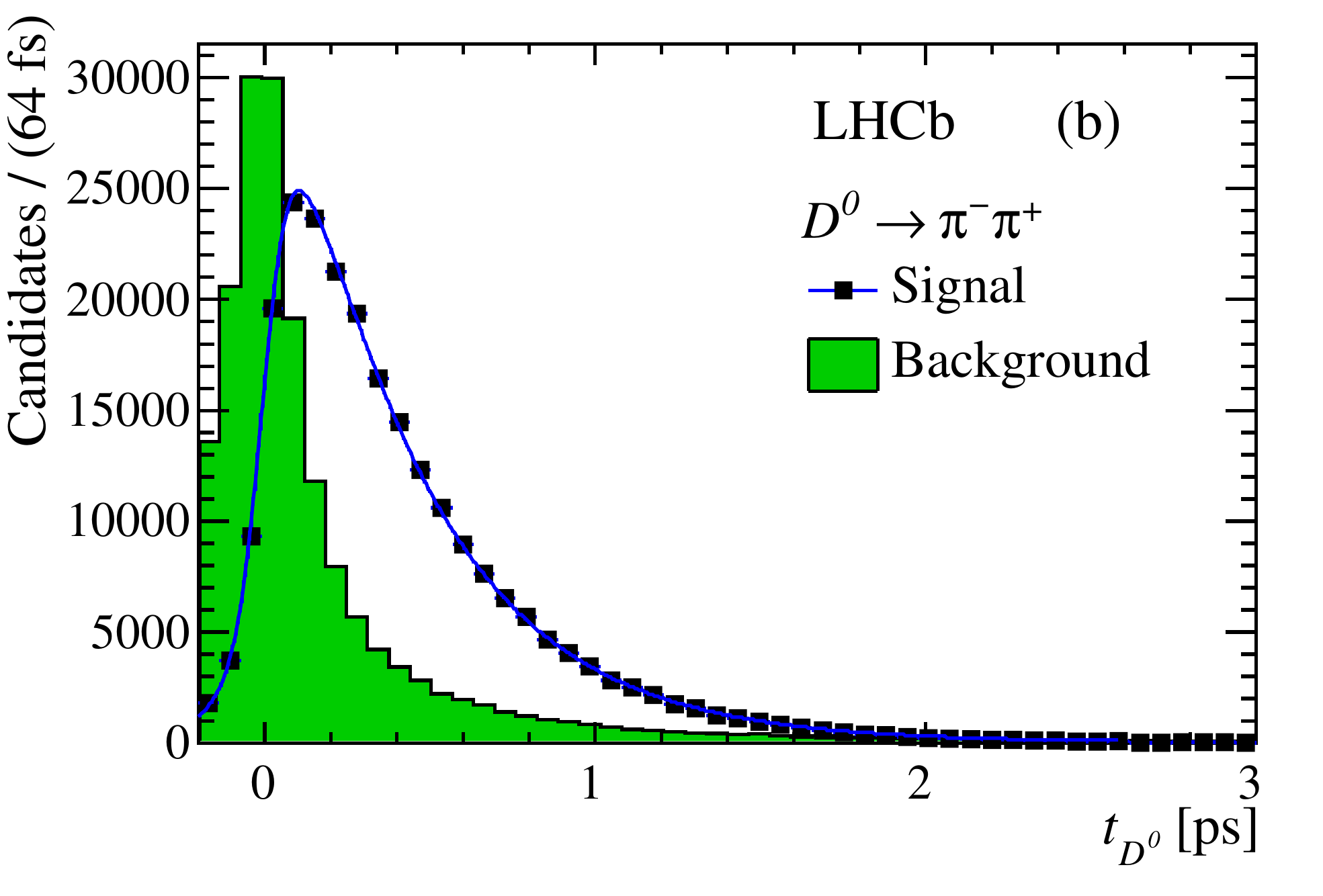}
  \end{center}
  \vspace{-0.95cm}
  \caption{\small Decay time distribution for signal candidates (solid points)
    with the result from the fit overlaid for (a) \dkk and (b) \dpipi
    decays. The distribution for background candidates scaled to a $\pm34\mevcc$
    window around the nominal \Dz mass is shown in the shaded (green)
    region. The distributions for signal and background candidates are obtained
    using the \sPlot~method.}
  \label{fig:lifetime_fits}
\end{figure}

The determination of the mean decay time is performed through a fit to the decay
time distribution of the signal candidates. Candidates with negative measured
decay times are included in the fit to have a better handle on the acceptance
and the resolution function. The measured decay time distribution is modelled by
a decreasing exponential function, with mean lifetime $\tau$, convolved with a
double Gaussian resolution function and multiplied with an acceptance function
of the form
\begin{equation}
  A(t) = 1 - a e^{-(t/(b\tau))^2} \ ,
\end{equation}
where $a$ and $b$ are acceptance parameters. The fit model is motivated by
simulation studies. The values for the fraction and width of the second Gaussian
and the acceptance parameter $b$ are taken from the simulation and fixed in the
fit. The 
role of the acceptance parametrisation is to allow a fit to the distribution
such that the resolution effect can be removed and the true decay time, which
appears in Eq.~(\ref{eq:AcpDirectIndirect}), can be evaluated. The observed
decay time distributions with the fit result superimposed are shown in
Fig.~\ref{fig:lifetime_fits}.

The decay time resolutions obtained from the lifetime fit (taken as the width of
the first Gaussian function) are $63.3\pm0.3\fs$ for \dkk and $58.3\pm0.4\fs$
for \dpipi, which are about 10\% larger than expected from simulations. The main
systematic uncertainties come from the uncertainty in the acceptance function
and from backgrounds. Using the world average of the \Dz lifetime,
$\tau(\Dz)=410.1\pm 1.5\fs$, the difference and average of the mean decay times
relative to $\tau(\Dz)$ are found to be
\begin{align}
  \Delta\mean{t}/\tau(\Dz) & =   0.018\pm0.002\stat\pm0.007\syst \nonumber\\
  \overline{\mean{t}}/\tau(\Dz) & =  1.062\pm0.001\stat\pm0.003\syst
  \ , \nonumber
\end{align}
where the uncertainty in $\tau(\Dz)$ is included as a systematic uncertainty.
Note that $\overline{\mean{t}}$ is not a measurement of the \Dz effective
lifetime (\ie, the lifetime measured with a single exponential fit), since this
number contains effects from the LHCb acceptance. The small value of
$\Delta\mean{t}$ implies that the measured value of \DACP is equal to the
difference in direct \CP violation, \ie, $\DACP=\Delta a_{\CP}^{\rm dir}$ with
negligible corrections.

\section{Systematic uncertainties}
\label{sec:systematics}

The contributions to the systematic uncertainty on \deltaACP are described
below.

\begin{itemize}
\item {\bf Difference in {\boldmath \bquark}-hadron mixture.} Due to the
  momentum requirements in the trigger and selection, the relative contribution
  from \Bz and \Bp decays (the contribution from \bquark-baryon and \Bs decays
  can be neglected) can be different between the \dkk and \dpipi modes. In
  combination with a different effective production asymmetry for candidates
  from \Bz and \Bp mesons (the production asymmetry from \Bz mesons is diluted
  due to \Bz mixing) this could lead to a non-vanishing bias in \DACP. Assuming
  isospin symmetry, the production cross-sections for \Bz and \Bp mesons are
  expected to be equal. Therefore, the ratio between \Bz and \Bp decays is
  primarily determined by their branching fractions to the $\Dz\Pmu\PX$ final
  state. Using the inclusive branching fractions~\cite{PDG2012},
  $\B^{+,0}\to\Dzb\PX$, the \Bz fraction is expected to be
  $f(\Bz)=(37.5\pm2.9)\%$. From the simulation the difference in the \Bz
  fraction due to the difference in selection efficiencies is found to be at
  maximum $1\%$. Further assuming a \Bp production asymmetry of
  $1.0\%$~\cite{LHCB-PAPER-2011-029} and assuming no \Bz production asymmetry,
  the difference in the effective production asymmetry between the two modes is
  $\sim0.02\%$.
\item{\bf Difference in {\boldmath \B} decay time acceptance.}  A difference
  between the \dkk and \dpipi modes in the \B decay time acceptance, in
  combination with \Bz mixing, changes the effective \B production
  asymmetry. Its effect is estimated from integrating the expected \B decay time
  distributions at different starting values, such that the mean lifetime ratio
  corresponds to the observed \PB decay length difference ($\sim5\%$) in the two
  modes. Using the estimated \Bz fraction and assuming a $1.0\%$ production
  asymmetry, the effect on \DACP is found to be $0.02\%$.
\item {\bf Effect of the weighting procedure.} After weighting the \Dz
  distributions in \pt and $\eta$, only small differences remain in the muon
  kinematic distributions. In order to estimate the systematic uncertainty from
  the \B production and detection asymmetry due to residual differences in the
  muon kinematic distributions, an additional weight is applied according to the
  muon $(\pt,\eta)$ and the azimuthal angle \mphi. The value of \DACP changes by
  $0.05\%$.
\item{\bf Difference in mistag asymmetry.}  The difference in the mistag rate
  between positive and negative tags contributes to the measured raw
  asymmetry. The mistag difference using \dkpi decays is measured to be
  $\Delta\mistag=(0.006\pm0.021)\%$ (see Sect.~\ref{sec:mistag}).  In case
  $\Delta\mistag$ is different for \dkk and \dpipi there can be a small effect
  from the mistag asymmetry. A systematic uncertainty of $0.02\%$ is assigned,
  coming from the uncertainty on $\Delta\mistag$.
\item{\bf Effect of different fit models.} A possible asymmetry in the
  background from false \Dz combinations is accounted for in the fit to the \Dz
  mass distribution. Different models can change the fraction between signal and
  background and therefore change the observed asymmetry. The baseline model is
  modified by either using a single Gaussian function for the signal, a single
  Gaussian plus a Crystal Ball function for the signal, a first- or second-order
  polynomial for the background, by leaving the asymmetry in the reflection
  free, or by modifying the fit range for \dpipi to exclude the reflection peak.
  The largest variation changes the value of \DACP by $0.035\%$. As another
  check, the asymmetry is determined without any fit by counting the number of
  positively- and negatively-tagged events in the signal window and subtracting
  the corresponding numbers in the sideband windows. The sideband windows are
  defined as $[\mu_{\rm sig}-48\mevcc,\mu_{\rm sig}-34\mevcc]$ and $[\mu_{\rm
      sig}+34\mevcc,\mu_{\rm sig}+48\mevcc]$, and the signal window as
  $[\mu_{\rm sig}-14\mevcc,\mu_{\rm sig}+14\mevcc]$, where $\mu_{\rm sig}$ is
  the mean of the signal distribution. This method changes the value of \DACP by
  $0.05\%$, which is taken as a systematic uncertainty.
\item{\bf Low-lifetime background in {\boldmath \dpipi}.}  As can be seen in
  Fig.~\ref{fig:lifetime_fits}, there is more background around $t=0$ in the
  \dpipi channel compared to the \dkk channel. If this background exhibits a
  non-flat or peaking structure this could bias the measurement of \DACP. When
  including the negative lifetime events the value of \DACP changes by $0.11\%$.
  This shift is taken as a systematic uncertainty.
\item{\bf {\boldmath \Lc} background in {\boldmath \dkk}.} A non-negligible
  fraction of the background in the \dkk mode originates from partial
  reconstruction of $\L_c^+\to\proton\Km\pip$ decays, where the proton is
  misidentified as a kaon. Most of these \Lc decays are expected to come from
  semileptonic \Lb decays. From exclusively reconstructed \Lc decays the shape
  of the background is observed to be linear in the $\Km\Kp$ invariant mass
  distribution. The influence of such a linear background on the fit model
  is tested by generating many pseudo-experiments. With an asymmetry in the \Lc
  background of $3\%$, which is a conservative upper bound for the asymmetry
  observed in the exclusively reconstructed $\Lc$ decays, a small bias of
  $0.03\%$ is seen in the measured asymmetry. This bias is taken as a systematic
  uncertainty.
\end{itemize}

The systematic uncertainties are summarised in Table~\ref{tab:sys_overall}. The
effects from higher-order corrections to Eq.~(\ref{eq:Araw}) and of the
uncertainty in the average mistag rate are found to be negligible. The overall
systematic uncertainty on \DACP, obtained by adding the individual contributions
in quadrature, is $0.14\%$.

\begin{table}
  \begin{center}
    \caption {\small Contributions to
    the systematic uncertainty of \DACP.}
    \vspace{0.1cm}
    \begin{tabular}{l|c}
                                                & Absolute    \\
      Source of uncertainty                     & uncertainty \\ \hline
      {\bf Production asymmetry:}               &             \\
      ~~~Difference in \bquark-hadron mixture   & $0.02\%$    \\
      ~~~Difference in \B decay time acceptance & $0.02\%$    \\
      {\bf Production and detection asymmetry:} &             \\
      ~~~Different weighting                    & $0.05\%$    \\
      {\bf Background from real \boldmath{\Dz} mesons:}  &    \\
      ~~~Mistag asymmetry                       & $0.02\%$    \\
      {\bf Background from fake \boldmath{\Dz} mesons:}  &    \\
      ~~~\Dz mass fit model                     & $0.05\%$    \\
      ~~~Low-lifetime background in \dpipi      & $0.11\%$    \\
      ~~~\Lc background in \dkk                 & $0.03\%$    \\ \hline
      {\bf Quadratic sum}                       & $0.14\%$    \\
    \end{tabular}
  \end{center}
  \label{tab:sys_overall}
\end{table}

\section{Cross-checks}
\label{sec:crosschecks}

\begin{figure}
  \begin{center}
    \includegraphics[width=0.49\textwidth]{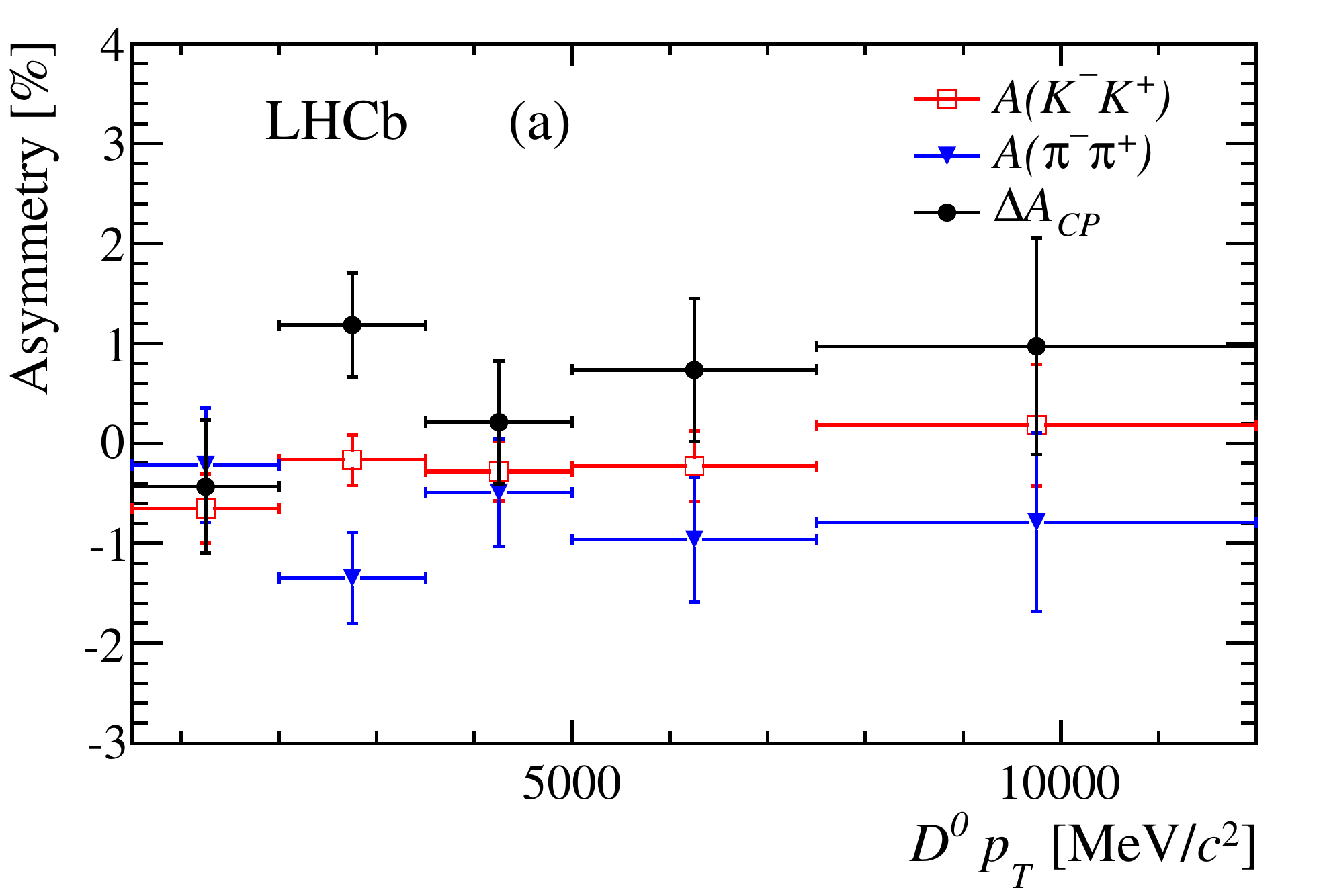}
    \includegraphics[width=0.49\textwidth]{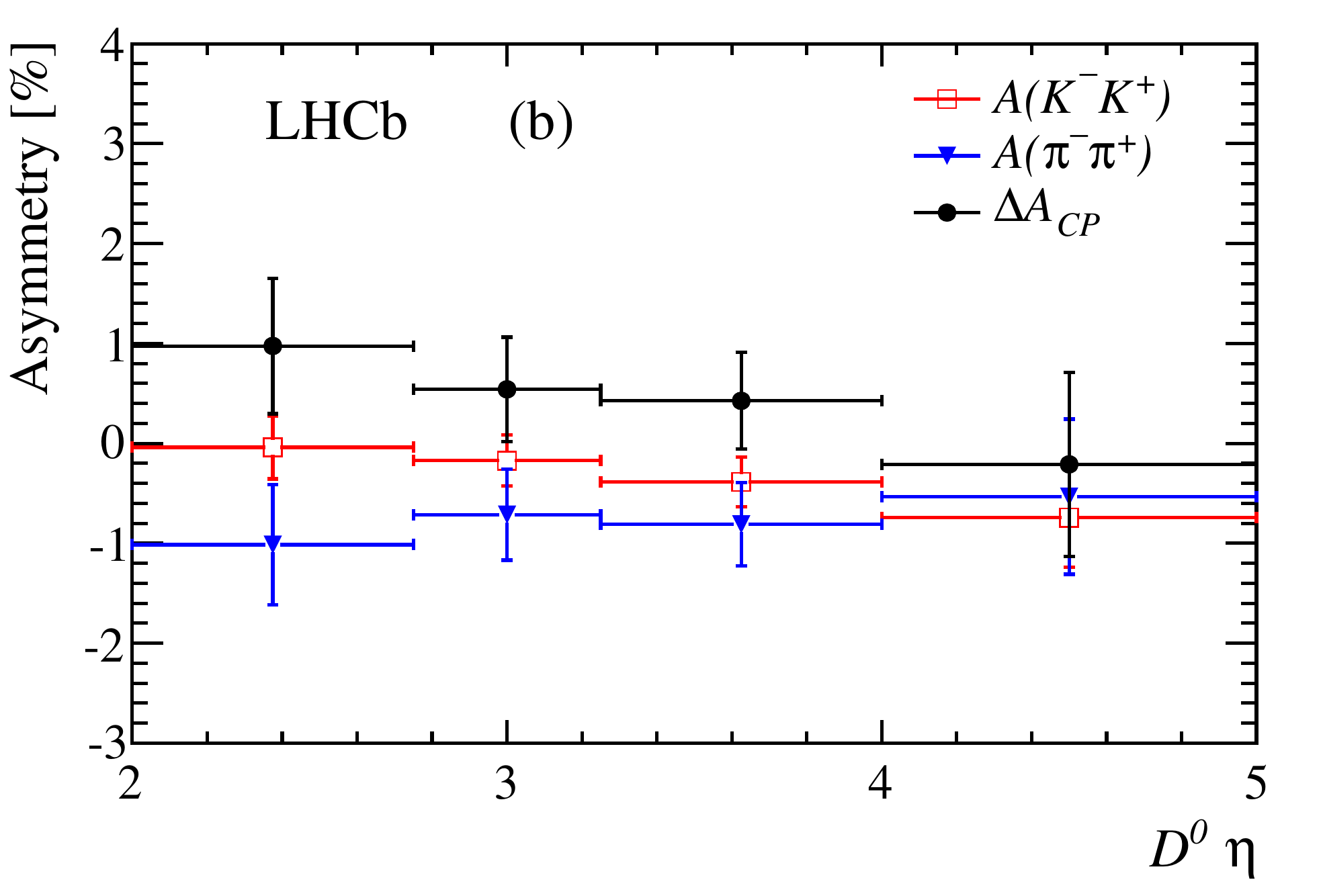}
    \end{center}
    \vspace*{-1.0cm}
    \caption{\small Raw asymmetries and \DACP as a function of (a) \pt and (b)
      $\eta$ of the \Dz meson. No weighting is applied.}
    \label{fig:crosscheck_d0}
\end{figure}

Many cross-checks have been performed to verify the stability of the result. In
particular, the raw asymmetries and \DACP are found to be stable when applying
fiducial cuts in the two-dimensional space of the muon momentum and its
horizontal component, when comparing different trigger decisions and when
applying tighter particle identification requirements on the \Dz daughters or on
the muons. The stability of the raw asymmetries and \DACP is also investigated
as a function of all possible reconstructed quantities, for instance the \Dz
decay time, the \bquark-hadron flight distance, the reconstructed \Dz--muon
mass, the angle between the muon and \Dz daughters, and the (transverse) momenta
and pseudorapidity of the muon and \Dz meson. No significant dependence is
observed in any of these variables. For example, Fig.~\ref{fig:crosscheck_d0}
shows \DACP and the raw asymmetries in the \dkk and \dpipi modes as a function
of \pt and $\eta$ of the \Dz meson, which are the variables that are used in the
weighting procedure. To check for a possible time dependence of the detection
asymmetry the data taking period is divided into six parts of roughly equal
integrated luminosity. The six parts are separated by periods without beam and
changes in the magnet polarity. No significant variation of the raw asymmetries
is observed.

\section{Conclusion}

The difference in \CP asymmetries between the \dkk and \dpipi modes is measured
using \Dz mesons produced in semileptonic \PB decays and is found to be
\begin{equation}
   \DACP = (0.49\pm 0.30\stat \pm 0.14\syst)\% \ .
   \nonumber
\end{equation}
This result takes into account the muon mistag probability and differences in
the kinematic distributions of \dkk and \dpipi decays. When neglecting indirect
\CP violation the difference between this result and the previous published LHCb
result using prompt \Dz decays~\cite{LHCb-PAPER-2011-023} is $3.2$ standard
deviations, assuming that the uncertainties have a Gaussian distribution. The
discrepancy, however, is reduced to $2.2$ standard deviations comparing to a
preliminary update of the previous result~\cite{LHCb-CONF-2013-003}. This result
does not confirm the evidence for direct \CP violation in the charm sector.

\section*{Acknowledgements}

\noindent We express our gratitude to our colleagues in the CERN
accelerator departments for the excellent performance of the LHC. We
thank the technical and administrative staff at the LHCb
institutes. We acknowledge support from CERN and from the national
agencies: CAPES, CNPq, FAPERJ and FINEP (Brazil); NSFC (China);
CNRS/IN2P3 and Region Auvergne (France); BMBF, DFG, HGF and MPG
(Germany); SFI (Ireland); INFN (Italy); FOM and NWO (The Netherlands);
SCSR (Poland); ANCS/IFA (Romania); MinES, Rosatom, RFBR and NRC
``Kurchatov Institute'' (Russia); MinECo, XuntaGal and GENCAT (Spain);
SNSF and SER (Switzerland); NAS Ukraine (Ukraine); STFC (United
Kingdom); NSF (USA). We also acknowledge the support received from the
ERC under FP7. The Tier1 computing centres are supported by IN2P3
(France), KIT and BMBF (Germany), INFN (Italy), NWO and SURF (The
Netherlands), PIC (Spain), GridPP (United Kingdom). We are thankful
for the computing resources put at our disposal by Yandex LLC
(Russia), as well as to the communities behind the multiple open
source software packages that we depend on.


\addcontentsline{toc}{section}{References}
\bibliographystyle{LHCb}
\bibliography{main,LHCb-PAPER,LHCb-CONF,LHCb-DP}

\end{document}